\documentstyle[12pt,aaspp4]{article}

\begin{document}

%
\def\n{\footnotemark}
\def\IUE{{\it IUE}}
\def\HST{{\it HST}}
\def\ISO{{\it ISO}}
\def\deg{$^{\rm o}$}
\def\degC{$^{\rm o}$C}
\def\arcsec{\ifmmode '' \else $''$\fi}
\def\arcmin{$'$}
\def\arcsecpoint{\ifmmode ''\!. \else $''\!.$\fi}
\def\arcminpoint{$'\!.$}
\def\kms{\ifmmode {\rm km\ s}^{-1} \else km s$^{-1}$\fi}
\def\Msun{\ifmmode {\rm M}_{\odot} \else M$_{\odot}$\fi}
\def\Lsun{\ifmmode {\rm L}_{\odot} \else L$_{\odot}$\fi}
\def\Zsun{\ifmmode {\rm Z}_{\odot} \else Z$_{\odot}$\fi}
\def\ergsAcm{ergs\,s$^{-1}$\,cm$^{-2}$\,\AA$^{-1}$}
\def\ergscm2{ergs\,s$^{-1}$\,cm$^{-2}$}
\def\qo{\ifmmode q_{\rm o} \else $q_{\rm o}$\fi}
\def\Ho{\ifmmode H_{\rm o} \else $H_{\rm o}$\fi}
\def\ho{\ifmmode h_{\rm o} \else $h_{\rm o}$\fi}
\def\ltsim{\raisebox{-.5ex}{$\;\stackrel{<}{\sim}\;$}}
\def\gtsim{\raisebox{-.5ex}{$\;\stackrel{>}{\sim}\;$}}
\def\vFWHM{\ifmmode v_{\mbox{\tiny FWHM}} \else
            $v_{\mbox{\tiny FWHM}}$\fi}
\def\CCF{\ifmmode F_{\it CCF} \else $F_{\it CCF}$\fi}
\def\ACF{\ifmmode F_{\it ACF} \else $F_{\it ACF}$\fi}
\def\Halpha{\ifmmode {\rm H}\alpha \else H$\alpha$\fi}
\def\Hbeta{\ifmmode {\rm H}\beta \else H$\beta$\fi}
\def\Hgamma{\ifmmode {\rm H}\gamma \else H$\gamma$\fi}
\def\Hdelta{\ifmmode {\rm H}\delta \else H$\delta$\fi}
\def\Lya{\ifmmode {\rm Ly}\alpha \else Ly$\alpha$\fi}
\def\Lyb{\ifmmode {\rm Ly}\beta \else Ly$\beta$\fi}
\def\Lyg{\ifmmode {\rm Ly}\beta \else Ly$\gamma$\fi}
\def\hi{H\,{\sc i}}
\def\hii{H\,{\sc ii}}
\def\hei{He\,{\sc i}}
\def\heii{He\,{\sc ii}}
\def\ci{C\,{\sc i}}
\def\cii{C\,{\sc ii}}
\def\ciii{\ifmmode {\rm C}\,{\sc iii} \else C\,{\sc iii}\fi}
\def\civ{\ifmmode {\rm C}\,{\sc iv} \else C\,{\sc iv}\fi}
\def\ni{N\,{\sc i}}
\def\nii{N\,{\sc ii}}
\def\niii{N\,{\sc iii}}
\def\niv{N\,{\sc iv}}
\def\nv{N\,{\sc v}}
\def\oi{O\,{\sc i}}
\def\oii{O\,{\sc ii}}
\def\oiii{O\,{\sc iii}}
\def\o5007{[O\,{\sc iii}]\,$\lambda5007$}
\def\oiv{O\,{\sc iv}}
\def\ov{O\,{\sc v}}
\def\ovi{O\,{\sc vi}}
\def\neiii{Ne\,{\sc iii}}
\def\nev{Ne\,{\sc v}}
\def\neviii{Ne\,{\sc viii}}
\def\mgi{Mg\,{\sc i}}
\def\mgii{Mg\,{\sc ii}}
\def\mgx{Mg\,{\sc x}}
\def\siIV{Si\,{\sc iv}}
\def\siIII{Si\,{\sc iii}}
\def\siII{Si\,{\sc ii}}
\def\si{S\,{\sc i}}
\def\sii{S\,{\sc ii}}
\def\siii{S\,{\sc iii}}
\def\siv{S\,{\sc iv}}
\def\sv{S\,{\sc v}}
\def\svi{S\,{\sc vi}}
\def\caii{Ca\,{\sc ii}}
\def\feii{Fe\,{\sc ii}}
\def\feiii{Fe\,{\sc iii}}
\def\alii{Al\,{\sc ii}}
\def\aliii{Al\,{\sc iii}}
\def\piv{P\,{\sc iv}}
\def\pv{P\,{\sc v}}
\def\cliv{Cl\,{\sc iv}}
\def\clv{Cl\,{\sc v}}
\def\nai{Na\,{\sc i}}
\def\o{\o}
%

\title{An Atlas of Computed Equivalent Widths of \\Quasar Broad Emission Lines}

\author{Kirk Korista}
\affil{Department of Physics \& Astronomy, University of Kentucky,
Lexington, KY 40506}
\author{Jack Baldwin}
\affil{Cerro Tololo Interamerican Observatory\altaffilmark{1}, Casilla
603, La Serena, Chile}
\author{Gary Ferland \& Dima Verner}
\affil{Department of Physics \& Astronomy, University of Kentucky,
Lexington, KY 40506}

\altaffiltext{1}{Operated by the Association of Universities for
Research in Astronomy Inc.\ (AURA) under cooperative agreement with the
National Science Foundation}

\begin{abstract}
We present graphically the results of several thousand photoionization
calculations of broad emission line clouds in quasars, spanning seven
orders of magnitude in hydrogen ionizing flux and particle density. The
equivalent widths of 42 quasar emission lines are presented as contours
in the particle density -- ionizing flux plane for a typical incident
continuum shape, solar chemical abundances, and cloud column density of
$N(H) = 10^{23}$~cm$^{-2}$.  Results are similarly given for a small
subset of emission lines for two other column densities
($10^{22}$~cm$^{-2}$ and $10^{24}$~cm$^{-2}$), five other incident
continuum shapes, and a gas metallicity of 5~\Zsun\/.  These graphs
should prove useful in the analysis of quasar emission line data and in
the detailed modeling of quasar broad emission line regions. The
digital results of these emission line grids and many more are
available over the Internet.
\end{abstract}

\newpage
%
%
%
\section{Introduction}

Early photoionization calculations could roughly reproduce the average
observed quasar broad emission line (BEL) spectrum in a single
``cloud''; i.e.\ one with a single column density ($N(H) \sim
10^{23}$~cm$^{-2}$), gas density ($\sim 10^{10}$~cm$^{-3}$), and
ionization parameter ($U \equiv n_{ph}/n_{e} \sim 0.01$) (Davidson
1977, Davidson \& Netzer 1979, Kwan \& Krolik 1981). However, in the
past decade, spectroscopic observations have pointed to the presence of
a wide distribution in the emission line cloud properties within each
quasar (Gaskell 1982; Wilkes 1984; Espey et al.\ 1989; Corbin 1989;
Clavel et al.\ 1991; Peterson 1993), prompting the more recent
multi-cloud, single pressure law calculations by Rees, Netzer, \&
Ferland (1989), Krolik et al.\ (1991), Goad, O'Brien, \& Gondhalekar
(1993), and O'Brien, Goad, \& Gondhalekar (1994, 1995).  Most recently,
Baldwin et al.\ (1995) found that simply integrating the emission from
a wide distribution in cloud gas properties (gas density, ionizing
flux, and column density) results in a spectrum that is consistent with
composite quasar spectra and may offer simple solutions to some
observational problems (e.g., emission line profile differences and
reverberation). This hypothesis will be further investigated in future
papers (Korista et al.\ 1996).

In anticipation of future, more advanced modeling of the broad emission
line regions (BLRs) of quasars and active galactic nuclei (AGN), we
present here the results of thousands of photoionization calculations
that span seven orders of magnitude in hydrogen ionizing flux and
particle density. The equivalent widths of 42 of the more prominent
quasar emission lines are presented as contours in the particle density
-- ionizing flux plane, for a typical incident continuum shape, solar
chemical abundances, and cloud column density of $N(H) =
10^{23}$~cm$^{-2}$.  Results are similarly given for a small subset of
emission lines for two other column densities, five other incident
continuum shapes, and a gas metallicity of 5~\Zsun\/. We present this
as a foundation for future research, and also as an interpretational aid
for the spectroscopic observer. 

We describe the calculations and the emission line equivalent width
maps in $\S$~2 and conclude in $\S$~3. The digital results of these and
many more BEL photoionization grids will be available in electronic
form over the Internet. Instructions for electronic retrieval are also
given in $\S$~3.
  
\section{The Photoionization Grids}

\subsection{The Calculations}

The line spectrum emitted by individual quasar ``clouds'' (whatever
their nature) depends on a number of parameters, namely, the gas
density, the gas column density, the flux and shape of the incident
continuum, and the gas chemical abundances. The standard picture is
that each observed line is the result of a superposition of line
emission from a large number of clouds, each emitting at their thermal
width (Davidson \& Netzer 1979). The present calculations assume
thermal velocity line widths in each cloud. The presence of significant
turbulence or streaming motions within the emitting region would have
significant effects on the results presented below, and will be the
subject of future work.  These would mainly act to alter the line
escape probabilities (perhaps along preferred directions), thereby
reducing the effects of line thermalization at high density and optical
depth, and continuum pumping contributions to some emission lines could
also become significant.

After choosing a particular chemical abundance (initially solar),
incident continuum shape, and cloud hydrogen column density (initially
$10^{23}$~cm$^{-2}$), we computed an emission line spectrum for each
coordinate pair in the gas density -- hydrogen-ionizing photon flux
plane ($\log n(H)$, $\log \Phi(H)$) using the code {\sc Cloudy},
version 90.02d (Ferland 1996; Ferland et al.\ 1996). Constant hydrogen
density throughout each cloud was assumed.  The results tend to be
similar for a similar ionization parameter, defined $U(H) \equiv
\Phi(H)/n(H)c$; lines in the gas density -- ionizing flux plane with
45\deg\/ slopes will be of constant $U(H) \times c$. With an origin of
$\log n(H) = 7$, $\log \Phi(H) = 17$, the grid was stepped in 0.25 dex
increments, and spanned seven orders of magnitude in each direction,
for a total of 841 simulations per grid. This should more than cover
the parameter space of the broad emission line clouds. Gas densities $>
10^{14}$~cm$^{-3}$ were not considered, since in this regime the
calculations are not deemed sufficiently accurate and such clouds are
mainly continuum sources (Rees, Netzer, \& Ferland 1989). 

The chosen form of the incident continuum shape was a combination of a
$f_{\nu} \propto \nu^{-0.5} exp(-h\nu/kT_{cut}$) UV-bump with an X-ray
power law of the form $f_{\nu} \propto \nu^{-1}$ spanning 13.6~eV to
100~keV, appropriate for radio-quiet quasars and AGN (Elvis et
al.\ 1994).  We considered two values of the UV-bump cutoff
temperature, $T_{cut}$, such that the energy in the UV-bump peaked at
$E_{peak} \approx 22$~eV and 44~eV, corresponding to $\log T_{cut} =$
5.7, 6.0, respectively. The UV-bump was also cut off in the infrared
with a temperature $kT_{IR} = 0.01$~Ryd, corresponding to 9.1 microns.
Most of the observed IR continuum is thought to originate in warm dust,
far outside the BEL region (Sanders et al.\ 1989; Barvainis 1990,
1992).  Free-free heating from a strong incident IR continuum would
have profound effects on the ionization and thermal properties of high
density clouds (Ferland \& Persson 1989; Ferland et al.\ 1992).  The UV
and X-ray continuum components were combined using a range of UV to
X-ray logarithmic spectral slopes of $\alpha_{ox} = -1.2, -1.4$, and
$-1.6$ (note that we prefer the convention $f_{\nu} \propto
\nu^{\alpha}$). It is defined \begin{equation}
\frac{f_{\nu}(2~keV)}{f_{\nu}(2500~\AA\/)} = 403.3^{\alpha_{ox}}.
\end{equation} We plot the incident continua considered here in
Figure~1. For our baseline grid we chose the continuum which might
typify an average QSO:  $E_{peak} = 44$~eV and $\alpha_{ox} = -1.4$.
For the hardest value of $\alpha_{ox}$ considered here ($\alpha_{ox} =
-1.2$), generally applicable to Seyfert~1 type AGN, the slope of the
exponential portion of the UV-bump has a value of roughly $-$2.3,
similar to observed slopes of soft X-ray excesses (e.g., Walter \& Fink
1993).
 
\subsection{The Baseline Grid}

Our baseline grid assumes solar abundances, a cloud column density of
$10^{23}$~cm$^{-2}$, and the baseline incident continuum described
above. {\sc Cloudy} now considers the ionization balance of the first
thirty elements; solar abundances are from Grevesse \& Anders (1989)
and Grevesse \& Noels (1993).

\noindent
H :1.00E+00  He:1.00E-01  Li:2.04E-09  Be:2.63E-11  B :7.59E-10  
C :3.55E-04  N :9.33E-05  O :7.41E-04  F :3.02E-08 Ne:1.17E-04
Na:2.06E-06  Mg:3.80E-05  Al:2.95E-06  Si:3.55E-05  P :3.73E-07  
S :1.62E-05  Cl:1.88E-07  Ar:3.98E-06 K :1.35E-07  Ca:2.29E-06
Sc:1.58E-09  Ti:1.10E-07  V :1.05E-08  Cr:4.84E-07  Mn:3.42E-07
Fe:3.24E-05  Co:8.32E-08 Ni:1.76E-06  Cu:1.87E-08  Zn:4.52E-08

\subsubsection{The Electron Temperature at the Face of the Cloud}

Here we illustrate the dependence of the cloud gas temperatures on its
position in the gas density -- ionizing flux plane. In Figure~2 we plot
contours of equal electron temperature at the {\em face} ($T_{face}$)
of the cloud in this plane, assuming solar chemical abundances and the
baseline incident continuum ($E_{peak} = 44$~eV and $\alpha_{ox} =
-1.4$). Both parameters will affect the details of this plot, but the
trends will remain the same. This temperature is proportional to the
ionization parameter, being $\sim 7000$~K in the lower right corner and
increasing slowly at first with increasing ionization parameter until
$\log U(H) \sim -1$. Line thermalization is not important at the
illuminated face of the cloud, thus its temperature is not strongly
dependent upon the gas density at constant $U(H)$. However, other
cooling and heating mechanisms do change with density along constant
$U(H)$, and this accounts for the generally gentle bending of the
contours with increasing density (the gas is hotter at higher density
along constant ionization parameter). In particular, free-free heating
becomes very important at high densities, and will depend upon the
shape and intensity of the incident infrared continuum. With increasing
$U(H)$ the temperature at the face of the cloud climbs more rapidly
than at low $U(H)$, reaching $10^5$~K at $\log U(H) \sim 1$.  Gas near
this temperature is cooling mainly via \ovi\/ $\lambda$1034 and then
\neviii\/ $\lambda$774, with a diminishing contribution from \civ\/
$\lambda$1549.  Once these ions begin to disappear, the gas can no
longer cool efficiently; the temperature contours become very dense
just beyond $\log U(H) \sim 1$, and the gas temperature jumps from
$\sim 10^5$~K to $\sim 10^6$~K; this is the well known thermal
instability gap (see for instance Krolik, McKee, \& Tarter 1981;
Reynolds \& Fabian 1995). It is this feature which will be particularly
sensitive to continuum shape and chemical abundances assumed. For
example, the band of dense contours lying between $\log T = 5$ and 6,
representing the instability gap, moves to lower ionization parameter
for harder continua. This thermally unstable gas or gas just a bit
cooler (and thermally stable) may be present near the central engines
of AGN. Gas with temperatures of $1-2 \times 10^5$~K and column
densities of $\sim 10^{22}$~cm$^{-2}$ has been observed in AGN soft
X-ray spectra in the form of ``warm absorbers'' (Fabian et al.\ 1994;
George, Turner, \& Netzer 1995). The identifications of significant
\neviii\/ $\lambda$774 emission in some QSO spectra (Hamann et
al.\ 1995; Hamann, Zuo, \& Tytler 1995) indicates the presence of the
same gas in higher column density clouds ($>10^{23}$~cm$^{-2}$).
This ``warm'' phase gas warrants further study. At still larger
ionization parameters the gas temperature climbs toward the Compton
temperature for the assumed incident continuum ($4.7 \times 10^6$~K).
The temperature again becomes insensitive to further increases in the
ionization parameter, as Compton heating and cooling are in balance.

\subsubsection{Emission Line Equivalent Widths} 

In Figures~3a -- 3g we present contours of logarithmic equivalent width
of 42 quasar emission lines in the $\log n(H)$ -- $\log \Phi(H)$
plane.  The equivalent width ($W_{\lambda}$ in \AA\/) is measured
relative to the incident continuum at 1216~\AA\/ for full source
coverage, and is a measure of the cloud's ability to reprocess the
continuum into the emission line. Here, full source coverage means that
the ``sky'', from the perspective of the continuum source, is
completely covered by clouds: $f_c \equiv 1$.  Results for cloud
covering fractions less than 1 may be obtained through a simple
rescaling. The $W_{\lambda}$ distributions are laid out in order of
increasing wavelength of the emission line, and the minimum contour
plotted has a value of 1~\AA\/ in every case.  Most of the first seven
far-UV emission lines have rarely, if ever, been reported, and are
plotted in light of the recent {\em Hubble Space Telescope} spectra of
$z \sim 1$ quasars.  Baldwin et al.\ (1995) discussed some of the
details of the $W_{\lambda}$ distributions, and we elaborate on some of
their discussion below.

\subsubsection{The Collisionally-Excited Metal Lines}

Collisionally excited lines such as \civ\/ $\lambda$1549 (Figure~3d)
generally show a band of efficient reprocessing running at constant
$U(H) \equiv \Phi(H)/n(H)c$ along the {\em center} of a diagonal ridge
from high $\Phi(H)$ and $n(H)$, to low $\Phi(H)$ and $n(H)$. For \civ\/
$\lambda$1549 this ridge corresponds to a $\log U(H) \approx -1.5$. The
gentle decrease in $W_{\lambda}$ along the ridge at constant $U(H)$ is
the result of thermalization, described further below. The
$W_{\lambda}$ decreases sharply when moving orthogonal to the ridge
because $U(H)$ is either too low (lower right) or too high (upper left)
to produce the line efficiently. For a column density of
$10^{23}$~cm$^{-2}$, clouds with $\log U(H) \gtsim 0.5$ are optically
thin to He$^+$ ionizing photons ($h\nu \gtsim 54$~eV) and so reprocess
little of the incident continuum. \civ\/ $\lambda$1549 can be
contrasted with a lower ionization line such as \ciii\/ $\lambda$977
(Figure~3b), whose peak $W_{\lambda}$ is shifted to lower $\log U(H)$
($\approx -2.25$), or to the high ionization \ovi\/ $\lambda$1034
(Figure~3b), shifted to higher $\log U(H)$ ($\approx 0$).  

A few effects modify the ridge of peak emission efficiency.  The high
ionization potential, high excitation energy of the \neviii\/
$\lambda$774 emission line along with the finite cloud column density
together conspire to make its ridge of reprocessing efficiency very
narrow in Figure~3a. The required ionization parameter is large, $\log
U(H) > 0.5$, and clouds with column densities of $10^{23}$~cm$^{-2}$
are becoming optically thin to He$^+$ ionizing photons; this line is
nearly fully formed within such clouds (see Hamann et al.\ 1995).
However, to fully form the peak ridge in the $W_{\lambda}$ distribution
of \neviii\/ $\lambda$774 in the density -- ionizing flux plane
requires larger column densities and ionization parameters ($\log
N(H)~\sim 10^{24}$~cm$^{-2}$, $\log U(H)~\sim~1$). Because of their
narrowness, the $W_{\lambda}$(\neviii\/) contours are best viewed by
looking obliquely along the diagonal ridge.  This effect will also be
important to \mgx\/ $\lambda$615, and to a lesser degree \ovi\/
$\lambda$1034, at this cloud column density.

The second effect is thermalization. This is most readily seen when
comparing an intercombination line with a resonance line of the same
ion. The classic example is \ciii\/ $\lambda$977 and \ciii\/]
$\lambda$1909 (Figures~3b and 3e).  While their ridges of peak
$W_{\lambda}$ have nearly the same ionization parameter ($\log U(H)
\approx -2.5$), the intercombination line begins to thermalize at
densities $\ltsim 10^{9.5}$~cm$^{-3}$, while the resonance line does
not begin to thermalize until the density has reached $\gtsim
10^{12}$~cm$^{-3}$. A smaller gas density at the same $U(H)$ means
smaller ionizing fluxes, corresponding to larger distances from an
isotropically emitting continuum source. These two lines may be emitted
in very different clouds. Thermalization at large optical depth is the
reason for the peculiar $W_{\lambda}$ distributions in lines such as
\oi\/ $\lambda$1304 and \mgii\/ $\lambda$2798 (Figures~3c and 3f). The
\oi\/ line is also extremely sensitive to the presence of the hydrogen
ionization front within the cloud; this accounts for the ``noise'' in
the contours along its upper boundary. Emission lines which continue to
become stronger at densities $> 10^{11}$~cm$^{-3}$, as the traditional
coolants become thermalized, are those which lie in the Lyman
continuum, \ciii\/ $\lambda$977, \niii\/ $\lambda$990, \ciii\/$^*$
$\lambda$1176, \nv\/ $\lambda$1240, \cii\/ $\lambda$1335, \siIV\/
$\lambda$1397, \aliii\/ $\lambda$1859, and \mgii\/ $\lambda$2798. In
the optical, \caii\/ (H \& K and the near-IR triplet), and \nai\/
$\lambda$5895, become increasingly strong at very high densities. The
strengths of these lines should be important indicators of the
importance of ionized high density clouds in the BLR.

One other effect is readily apparent in the $W_{\lambda}$ distributions
of those lines formed in the He$^+$ zone: in the gas density --
ionizing flux plane a secondary, more localized ridge of high
efficiency forms at larger $U(H)$ for gas densities $\gtsim
10^9$~cm$^{-3}$. This effect is seen as secondary peaks in the
$W_{\lambda}$(\niii\/] $\lambda$1750) and $W_{\lambda}$(\ciii\/]
$\lambda$1909) located near the coordinates (10.00,19.75; Figure~3e)
and (9.50,19.50; Figure~3e), respectively, or as bulges in the
$W_{\lambda}$ distributions of \cii\/ $\lambda$1335, \ciii\/
$\lambda$977, \niii\/ $\lambda$990, \oiii\/] $\lambda$1663, and
\siIV\/ $\lambda$1397. For the relevant ionization parameters, an
extended H$^+$ -- He$^+$ Str\"{o}mgren zone forms in these clouds,
increasing the relative emitting volumes for the aforementioned lines.
This line emission eventually thermalizes at higher flux levels at
constant $U(H)$ (i.e., for $\log n(H) \gtsim 10^{10-11}$~cm$^{-3}$).
This effect will be more important for higher column density clouds.
Note that considerable overlap in the efficient emission of \ciii\/
$\lambda$977 and \ciii\/] $\lambda$1909 occurs for clouds whose
properties lie in this region in the gas density -- ionizing flux
plane, in contrast to the ridge of primary efficiency at lower $U(H)$,
as discussed above.

It should be noted by the reader that dielectronic recombination
contributes significantly to the emission of several high excitation
lines, such as \cii\/ $\lambda$1335, \ciii\/ $\lambda$977, \ciii\/$^*$
$\lambda$1176, \niii\/ $\lambda$990, \niv\/ $\lambda$764, \oiii\/
$\lambda$835, and \ov\/ $\lambda$630. This contribution tends to be
more important for the lower ionization lines, because the collisional
rates are small due to the Boltzmann factor.  The recombination rates
were taken from Nussbaumer \& Storey (1983).  The recombination
contribution is also very important in the ``bulge'' and ``secondary
peak'' regions in the gas density -- ionizing flux plane for several of
the lines which have these features, just described above. This is
likely due to the fact that these lines at these positions in the gas
density -- ionizing flux plane are forming at larger depths in the
cloud, where the temperatures are lower.

Finally, we do not investigate here the complicated emission from
\feii\/.  The present version of {\sc Cloudy} predicts the heating and
cooling contributions from \feii\/ based upon a modified version of the
Wills, Netzer, \& Wills (1985) model, and will be discussed elsewhere
(Hamann et al.\ 1996). More advanced calculations of the emission of
\feii\/ and its role in the thermal balance of BEL clouds will be the
subject of future work (see Verner et al.\ 1995).

\subsubsection{The Hydrogen and Helium Lines}

Lines of H$^o$, He$^o$, and He$^+$ are largely produced via
recombination and emitted over a wider area on the gas density --
ionizing flux plane, including the low $\Phi(H)$ -- high $n(H)$
regions. This is because these ions still exist under these
conditions.  At sufficiently high ionization parameter for the given
cloud column density, the H, He, and He$^+$ ionization fronts reach the
backs of the clouds causing the dramatic declines in $W_{\lambda}$ of
the Balmer lines, \hei\/ $\lambda$5876, and \heii\/ $\lambda$4686 (see
Figures~3f, 3g).  These same lines have their peak efficiencies at high
densities ($\log n(H) \gtsim 12$), but different ionizing photon
fluxes. Note that in this high density regime, substantial emission
arises from collisional excitation. A feature common to most of the
hydrogen and helium emission lines is saturation at high continuum
fluxes. In contrast to the excited state transitions of H and He,
\Lya\/ $\lambda$1216 thermalizes at high density as well as high flux
(Figure~3c; see also \Lyb\/ an \Lyg\/ in Figure~3b).  Another common
feature is that they weaken in the direction of very high density and
very low flux (lower right corners) where either the ionized fractions
of the H, He ions are becoming very small or the temperature is
becoming too low to support the collisional excitation of the neutrals.
Note that the Balmer and Paschen continua ``$W_{\lambda}$''
distributions (Figures~3f and 3g) are the integrated fluxes from these
diffuse continua ratioed to the incident continuum flux-density at
1216~\AA\/.  Finally we note that the assumed ionizing spectrum
predicts a $W_{\lambda}(\Lya\/) \approx 1089$\AA\/ (3.04 in the
logarithm) for full coverage by an optically thick cloud if all
hydrogen ionizing photons are converted into \Lya\/. Compare this with
its computed $W_{\lambda}$ contours in Figure~3c. Thermalization and
destruction via background continuum opacities (Balmer continuum,
\feii\/, etc) act to diminish this conversion efficiency (Davidson \&
Netzer 1979; Shields \& Ferland 1993).

\subsection{Other Cloud Column Densities}

For solar abundances and the same incident continuum shape, Figures~4a
-- 4c show the contours of $\log W_{\lambda}$ for six selected emission
lines for cloud column densities of $10^{22}$~cm$^{-2}$,
$10^{23}$~cm$^{-2}$, and $10^{24}$~cm$^{-2}$, respectively. A
comparison between the two extremes, Figures~4a and 4c, shows the
expected results.  For the hydrogen and helium recombination lines
(left panels), the ionization parameter which fully ionizes the clouds
is simply proportional to the cloud column density: $N(H^+) \approx
10^{23.1} \times U(H)$~cm$^{-2}$.

This region of dramatic decline in $W_{\lambda}$ also shifts to lower
$U(H)$ with a decrease in the cloud column density for the
collisionally excited metal lines (right panels). The high ionization
lines, for example \ovi\/ $\lambda$1034, are affected in an additional
way:  the position of the ridge top shifts to lower $U(H)$ for smaller
column densities in order for a sufficient volume of O$^{+5}$ to exist
within the cloud. The combined effects of these act to first diminish
the lines' ridge $W_{\lambda}$ areas in the gas density -- ionizing
flux plane, and eventually to reduce their peak contour values as the
cloud column density is decreased. Thus, for a distribution in cloud
column densities the highest ionization lines are emitted most
efficiently in the highest column density clouds. As an illustration of
this we note that efficient emitters of \ovi\/ $\lambda$1034 are mainly
optically thick clouds whose column densities are $10^{24}$~cm$^{-2}$,
a roughly even mixture of optically thick/thin clouds whose column
densities are $10^{23}$~cm$^{-2}$, and mainly optically thin clouds
whose column densities are $10^{22}$~cm$^{-2}$. This can be seen by
comparing the ridge tops of this line for the three column densities to
a line representing the ionization parameter at which the cloud becomes
optically thin to hydrogen ionizing photons for that column density.

Finally, the $W_{\lambda}$ distributions in the low ionization lines,
such as \mgii\/ $\lambda$2798, primarily formed in a partially ionized
hydrogen zone (PIZ), are not strongly affected at significant
$W_{\lambda}$ until the cloud column density drops substantially below
$10^{22}$~cm$^{-2}$.  This is also true of the hydrogen and helium
emission lines at high density.

\subsection{Other Ionizing Continuum Shapes}

In Figures~5a -- 5e we show the $W_{\lambda}$ distributions for the
same six emission lines for solar abundances, a cloud column density of
$10^{23}$~cm$^{-2}$, and five other incident continuum shapes (see
Figure~1). The results for the baseline continuum are shown in
Figure~4b. Moving from the hardest (Figure~5a) to the softest
(Figure~5e) continuum, the major change for most emission lines is that
the value of the peak equivalent width diminishes, and/or the peak
contours shrink in area in the gas density -- ionizing flux plane.
When comparing the results between the two types of UV-bumps, about
0.15 dex of the changes in the peak line equivalent widths are due to
the fact that for the same number of hydrogen ionizing photons, a
softer UV bump will have more flux in the observed UV continuum upon
which the lines lie than a harder one (see Figure~1). The remaining
differences are due to the quantity of photons at relevant energies to
produce and excite the ions.  Some line ratios, such as \Lya\//\heii\/,
are expected to be sensitive mainly to the hardness of the UV-bump
($n_{ph}$(912~\AA\/)/$n_{ph}$(228~\AA\/)), while those involving the
strong metal line coolants (e.g., \Lya\//\civ\/ and \civ\//\heii\/)
should be dependent on both the hardness of the UV-bump and
$\alpha_{ox}$. It is also the case for many of the metal lines that the
upper boundary on $U(H)$ for significant reprocessing efficiency shifts
to lower $U(H)$ for harder continua.  This is because continua with
larger ratios of soft X-ray to 912~\AA\/ photons are more efficient in
stripping electrons from the ionized metals. This effect is especially
apparent in lines like \ovi\/ $\lambda$1034 (compare Figures~5a and
5e). Note, however, that the contours of lines formed in a PIZ, like
\mgii\/ $\lambda$2798, become larger in the directions of larger $U(H)$
for harder continua, because the volume of a PIZ within a cloud becomes
larger at constant $U(H)$ for harder continua (compare Figures~5a and
5e).

\subsection{Z$ = 5$~\Zsun\/}

Using the baseline incident continuum shape and a cloud column density
of $N(H) = 10^{23}$~cm$^{-2}$, we computed another grid assuming a
metallicity of 5~\Zsun\/, similar to that deduced for some
high-redshift quasars (Hamann \& Ferland 1993; Ferland et al.\ 1996).
We used the chemical abundance set derived by Hamann \& Ferland for
their model M5a and Z~$=$~5~\Zsun\/ (see Hamann \& Ferland and Ferland
et al.). The abundances, by number relative to hydrogen, are:

\noindent
H :1.00E+00  He:1.27E-01  Li:1.19E-08  Be:1.54E-10  B :4.44E-09  
C :8.06E-04  N :9.61E-04  O :5.18E-03  F :1.95E-07 Ne:7.55E-04
Na:1.36E-05  Mg:2.51E-04  Al:1.93E-05  Si:2.29E-04  P :2.41E-06  
S :1.06E-04  Cl:1.21E-06  Ar:2.57E-05 K :8.72E-07  Ca:1.42E-05
Sc:2.24E-09  Ti:1.56E-07  V :1.49E-08  Cr:6.87E-07  Mn:4.85E-07
Fe:4.60E-05  Co:1.18E-07 Ni:2.50E-06  Cu:2.65E-08  Zn:6.41E-08

\noindent
The $W_{\lambda}$ distributions for the same six selected emission
lines are shown in Figure~6. Comparing these results to those in
Figure~4b (same incident continuum and cloud column density, but solar
abundances), one can see several important differences. First, the
hydrogen and helium lines' peak $W_{\lambda}$ values are diminished at
a metallicity of 5~\Zsun\/ by a modest amount (0.1 -- 0.2 dex) compared
to solar.  This is because the metals become important opacity sources
with increasing metallicity, replacing to a significant degree,
hydrogen and helium.  Second, the peak values of the \ovi\/
$\lambda$1034 and \civ\/ $\lambda$1549 $W_{\lambda}$ distributions are
both diminished by about 0.4 dex. With increasing metallicity, Z, the
nitrogen abundance scales like Z$^2$ (Hamann \& Ferland 1993), and the
cooling from carbon and oxygen is shifting over to nitrogen. In
addition the strong metal opacity and accompanying lower electron
temperatures shrink the emitting volumes of especially the higher
ionization emission lines.  See Ferland et al.\ (1996) for specific
predictions regarding \nv\/, \heii\/ and Z-dependent behavior.

\subsection{Anisotropic Line Emission}
Davidson and Netzer (1979) and Ferland et al.\ (1992) discussed the
anisotropic line emission from clouds. O'Brien et al.\ (1994)
demonstrated the effects of this ``line beaming'' on the response
functions and line profiles of the emission lines. In general, strong
resonance lines, as well as optically thick excited state lines (e.g.,
\Hbeta\/ $\lambda$4861, \heii\/ $\lambda$1640) will generally
preferentially escape out the front sides of the clouds (i.e., the side
facing the incident continuum). The conditions which lead to
significant line beaming will differ for individual lines, depending
mainly on the ionization and temperature structure. For example, \Lya\/
becomes fully anisotropic with the formation of a hydrogen ionization
front within the cloud, but nearly isotropic in clouds which are
optically thin to hydrogen ionizing photons. Most other lines require
more specific conditions and are not so ``binary'' in their
inward/outward emission. In the case of \civ\/ $\lambda$1549, the
degree of the anisotropy is approximately proportional to the
ionization parameter. This is because the size of the ionized portion
of the cloud, and thus the C$^{3+}$ column density, is proportional to
$U(H)$. This line also becomes more anisotropic with increasing gas
density at constant ionization parameter, i.e., in the direction of
thermalization. The two effects result in anisotropic emission for the
following reason. Because the peak of the line source function within a
cloud does not coincide with the position of the peak ion fraction (the
former lies closer to the front face of the cloud where the temperature
is higher), the line is beamed toward the continuum source. In general
harder continua produce \civ\/ (and other lines) more isothermally in
this respect, and the line emission will be more isotropic than in
clouds which are illuminated by softer continua.

In Figure~7, we show contours of constant logarithmic ratios of
$F_{in}/F_{total}$ for \civ\/ $\lambda$1549, assuming the baseline
continuum, solar abundances, and cloud column densities of
$10^{23}$~cm$^{-2}$. The radiative transfer computations assume plane
parallel slab clouds; the actual line beaming function may be more
complicated depending on the cloud geometry, non-radial transfer, and
other effects. A comparison with Figure~4b shows that clouds should
emit \civ\/ anisotropically where its strength is significant. Low
density clouds which are becoming fully ionized in hydrogen are nearly
isotropic emitters (lower left corner), but are anisotropic emitters at
larger gas densities (moving toward the upper right corner). Clouds
optically thick to hydrogen ionizing photons beam \civ\/ preferentially
inward, with the beaming factor increasing as the line becomes
thermalized. The steep gradient in the inward/total ratio along the
upper diagonal boundary roughly coincides with the clouds becoming
optically thin to 54~eV photons, and thus with the steep decline in the
$W_{\lambda}$(\civ\/). This boundary shifts with cloud column density;
little else changes with changing column density in Figure~7. Other
incident continuum shapes, different chemical abundances, and the
presence of non-thermal motions would change the nature of the line
beaming significantly, however.

\section{Conclusions}
We have presented these grids to quantitatively show the general
dependence of the emission spectrum from individual BLR gas clouds to
changes in the most important input parameters. While most of these
trends will seem obvious in a qualitative sense, we hope that the
diagrams presented here will help future researchers to navigate more
easily through the $\Phi(H), n(H), N(H)$, continuum-shape, metallicity
parameter space.

The difference between the logarithmic equivalent width contour plots
for two different lines will be logarithmic contours of the line
intensity ratio, which is of considerable interest for analyzing
observational data. These grids should also prove valuable to 
the detailed modeling of broad line regions, since all such models
must include emissivity from a distribution in cloud properties.

We have chosen not to present intensity-ratio contour plots here (a
subject of future work), but rather to make the numerical data from the
equivalent width grids easily available over the Internet. These data
are accessible via the {\sc Cloudy} home page (currently
``http://www.pa.uky.edu/$\sim$gary/cloudy/''). A {\sc readme} file
describes the data files and will log their updates.  The data base
contains the information, as plotted in Figure~3, for over 100 emission
lines and diffuse continua in a 3 column ascii format file for each
emission line:  $\log n(H), \log \Phi(H), W_{\lambda1216}/1216$.
Information on how to convert a line's $W_{\lambda1216}/1216$ to it's
surface flux (useful for detailed modeling of the BLR) for a given
incident continuum shape are provided. All of these are available for a
variety of incident continuum shapes, many cloud column densities, a
few sets of gas abundances, and other parameters. 

We thank the referee, Fred Hamann, for his helpful comments and
suggestions.  This work was supported by the NSF (AST 93-19034), NASA
(NAGW-3315, NAG-3223), and STScI grant GO-2306.
%
%
%

%
\newpage
%
%
%
\begin{center}
{\bf Figure Captions}
\end{center}

\noindent
Fig. 1.--- {The continuum spectral energy distributions considered in
this paper. The UV-bump peaks at $E_{peak} \approx 44$~eV for the solid
lines and at $E_{peak} \approx 22$~eV for the dashed lines. Three
separate values of $\alpha_{ox}$ were considered, top to bottom: $-1.2,
-1.4, -1.6$. All continua plotted here have the same flux in hydrogen
ionizing photons.}

\noindent 
Fig. 2.--- {Contours of constant logarithm of the electron temperature
at the front face (first zone) of the cloud as a function of hydrogen
ionizing photon flux and hydrogen density. The baseline incident
continuum (a UV-bump that peaks near 44~eV and an $\alpha_{ox} =
-1.40$) and solar chemical abundances were assumed. The contours of
$10^4$~K and $10^6.6$~K are both labeled and increase monotonically
from the lower right corner to the upper left where the Compton
temperature for this continuum is reached ($4.7 \times 10^6$~K). The
solid contours are decades and the dotted are 0.1 dex increments.}

\noindent 
Fig. 3a.--- {Contours of $\log W_{\lambda}$ for six emission lines for
the baseline grid is shown as a function of the hydrogen density and
flux of hydrogen ionizing photons. The chemical abundances are solar,
the cloud column density is $10^{23}$~cm$^{-2}$, the continuum spectral
energy distribution has a UV-bump that peaks near 44~eV and an
$\alpha_{ox} = -1.40$. Line strengths are expressed as logarithmic
equivalent widths referenced to the incident continuum at 1216~\AA\/
for full source coverage. In this and in all figures that follow, the
smallest decade contoured is 1~\AA\/, each solid line is 1 dex, and
dotted lines represent 0.2 dex steps. In each case the peak in the
equivalent width distribution lies beneath the center of a solid
triangle. The contours generally decrease monotonically from the peak
to the 1~\AA\/ contour.  See text for further details. The solid star
is a reference point marking the ``standard BLR'' parameters discussed
by Davidson and Netzer (1979).}

\noindent
Fig. 3b.--- {Same as Figure~3a for other emission lines.}

\noindent
Fig. 3c.--- {Same as Figure~3a for other emission lines.}

\noindent
Fig. 3d.--- {Same as Figure~3a for other emission lines.}

\noindent
Fig. 3e.--- {Same as Figure~3a for other emission lines.}

\noindent
Fig. 3f.--- {Same as Figure~3a for other emission lines.}

\noindent
Fig. 3g.--- {Same as Figure~3a for other emission lines.}

\noindent
Fig. 4a.--- {Same as Figure~3 for a column density of $10^{22}$~cm$^{-2}$
and a selection of six emission lines.}

\noindent
Fig. 4b.--- {Same as Figure~3 for a column density of $10^{23}$~cm$^{-2}$
and a selection of six emission lines.}

\noindent
Fig. 4c.--- {Same as Figure~3 for a column density of $10^{24}$~cm$^{-2}$
and a selection of six emission lines.}

\noindent
Fig. 5a.--- {Same as Figure~4b for a continuum spectral energy
distribution whose UV-bump peaks at 44~eV and has an $\alpha_{ox} =
-1.20$.}

\noindent
Fig. 5b.--- {Same as Figure~4b for a continuum spectral energy
distribution whose UV-bump peaks at 22~eV and has an $\alpha_{ox} =
-1.20$.}

\noindent
Fig. 5c.--- {Same as Figure~4b for a continuum spectral energy
distribution whose UV-bump peaks at 22~eV and has an $\alpha_{ox} =
-1.40$.}

\noindent
Fig. 5d.--- {Same as Figure~4b for a continuum spectral energy
distribution whose UV-bump peaks at 44~eV and has an $\alpha_{ox} =
-1.60$.}

\noindent
Fig. 5e.--- {Same as Figure~4b for a continuum spectral energy
distribution whose UV-bump peaks at 22~eV and has an $\alpha_{ox} =
-1.60$.}

\noindent
Fig. 6.--- {Same as Figure~4b for enhanced metal abundances. A metallicity
of Z$ = 5$ was assumed, using the abundance grids of Hamann \& Ferland
(1993).}

\noindent
Fig. 7.---{Contours of logarithmic inward to total emission line flux
ratios for \civ\/ $\lambda$1549, assuming the baseline grid parameters.
The most important contours are labeled, with $-0.30$ designating
isotropic emission (inward/outward flux ratio$ = 1.0$), and $-0.05$
designating clouds emitting \civ\/ most anisotropically. The step size
for the contours is 0.05 dex. See text for details.
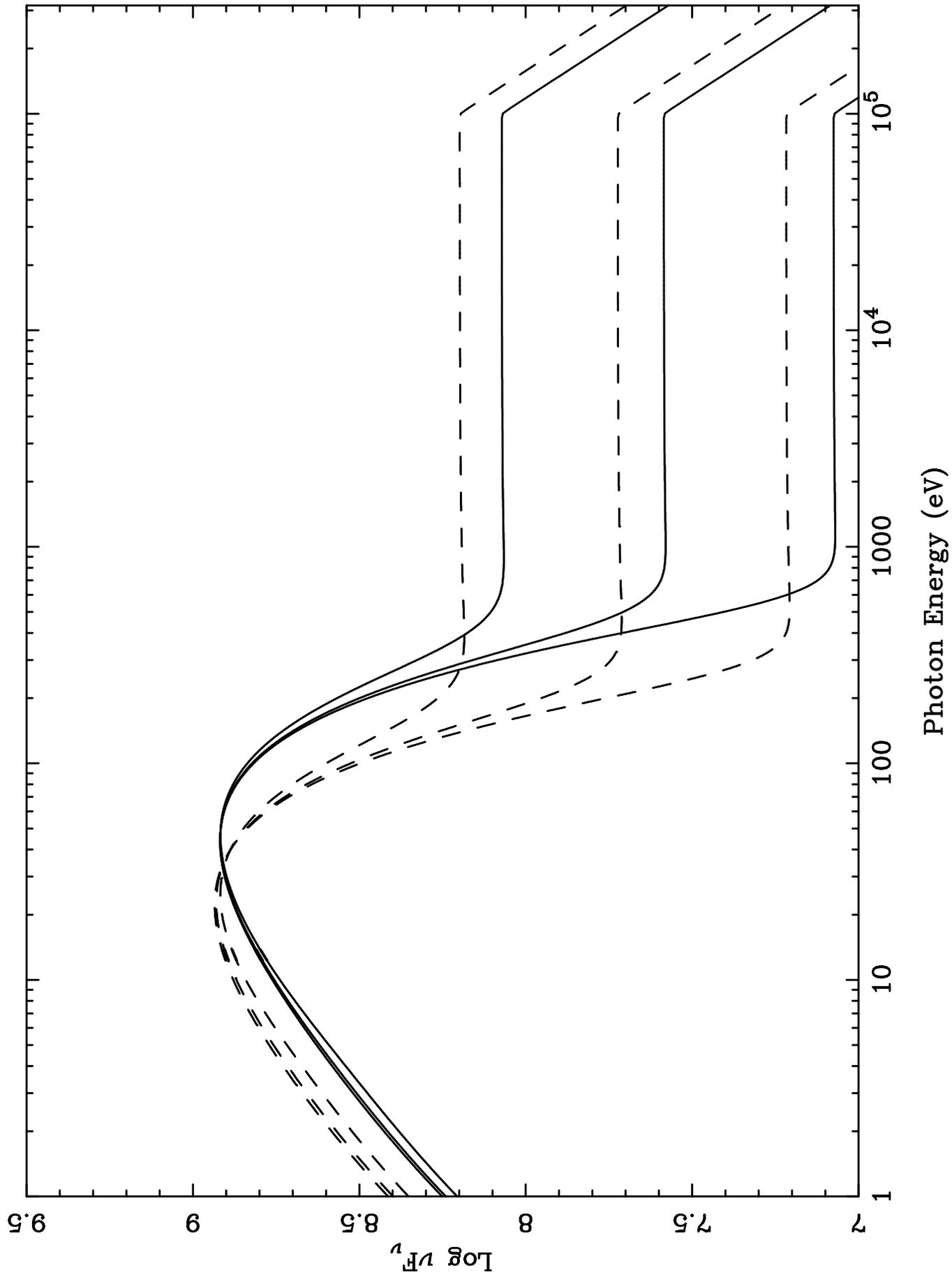
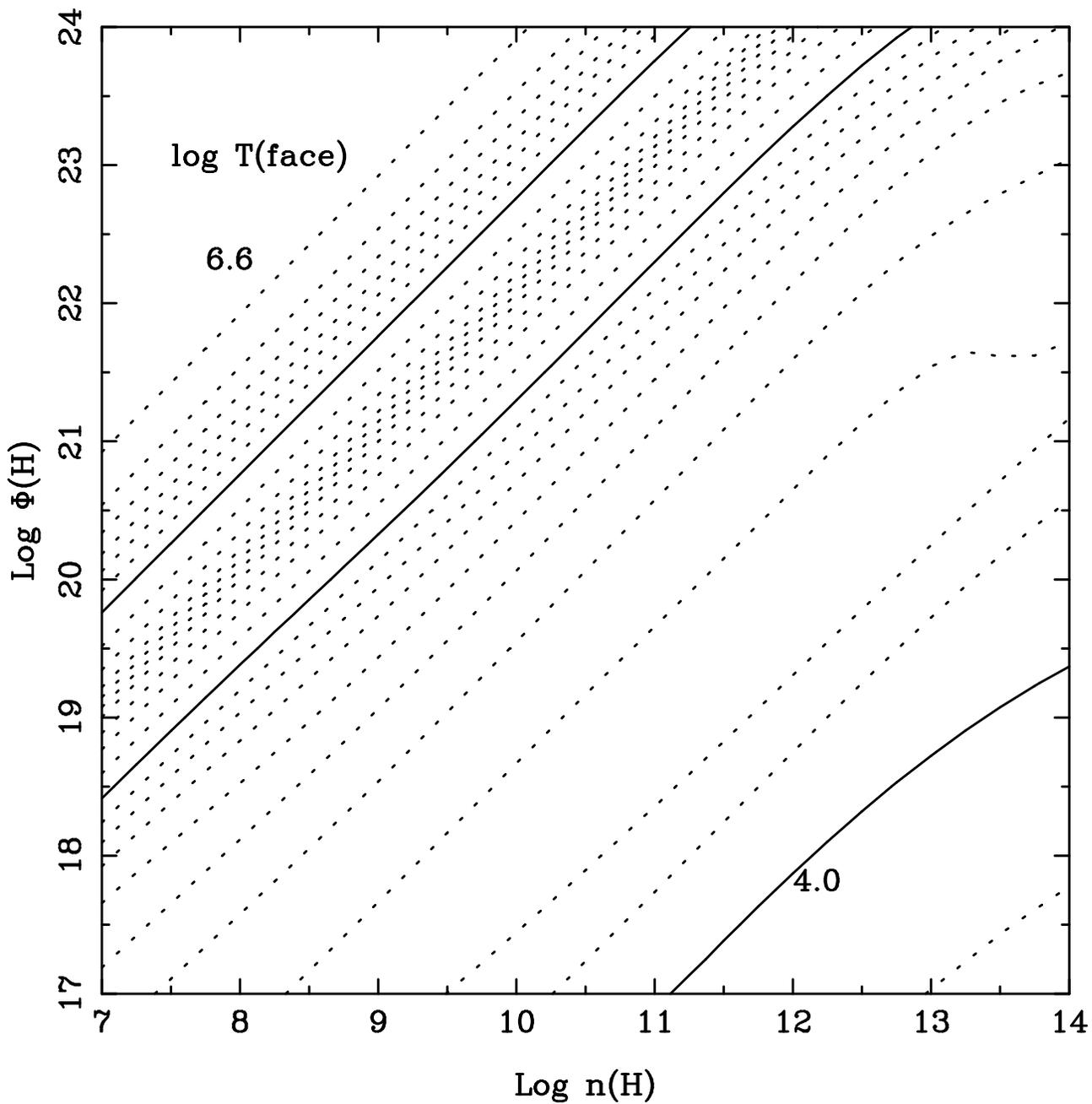
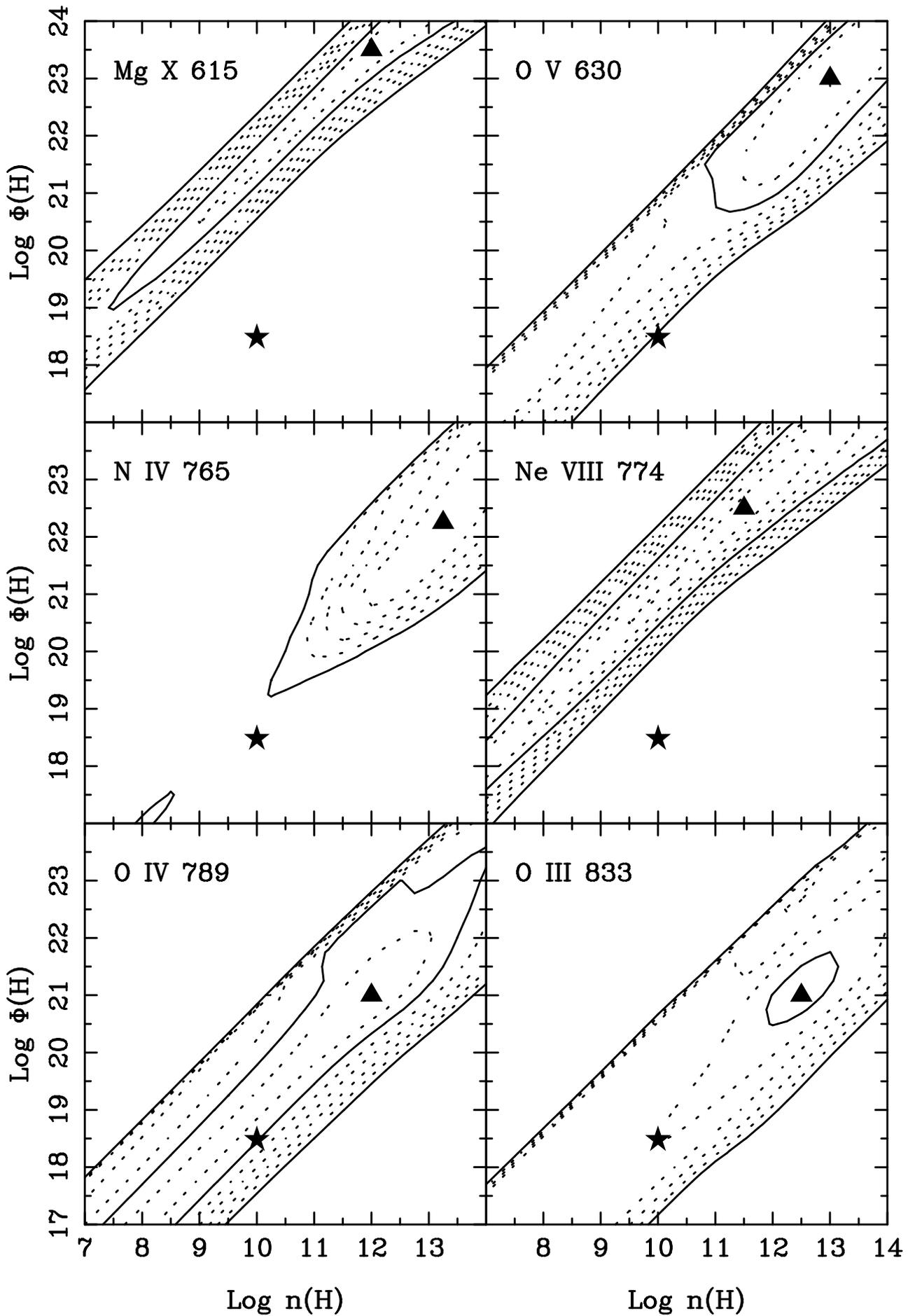
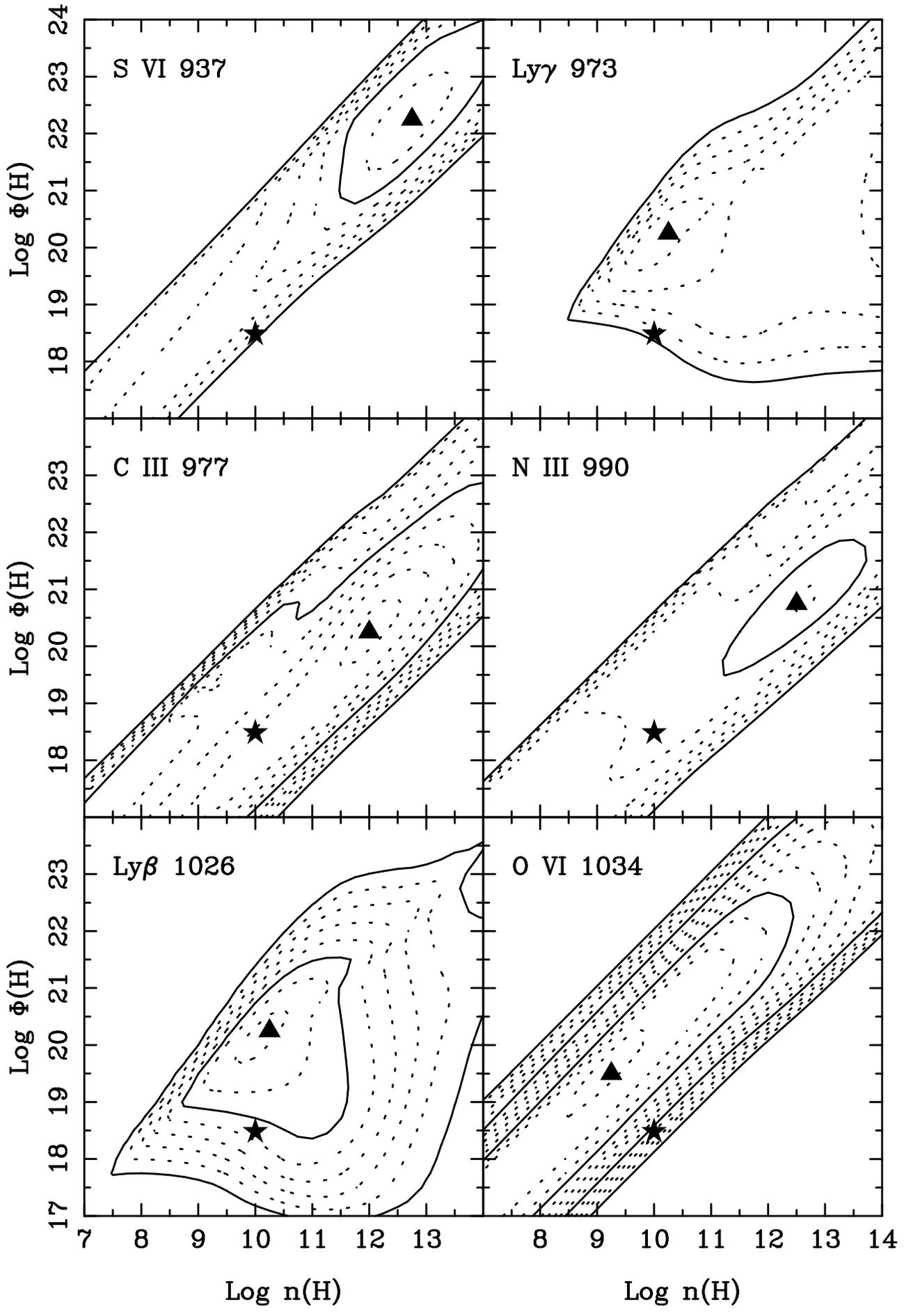
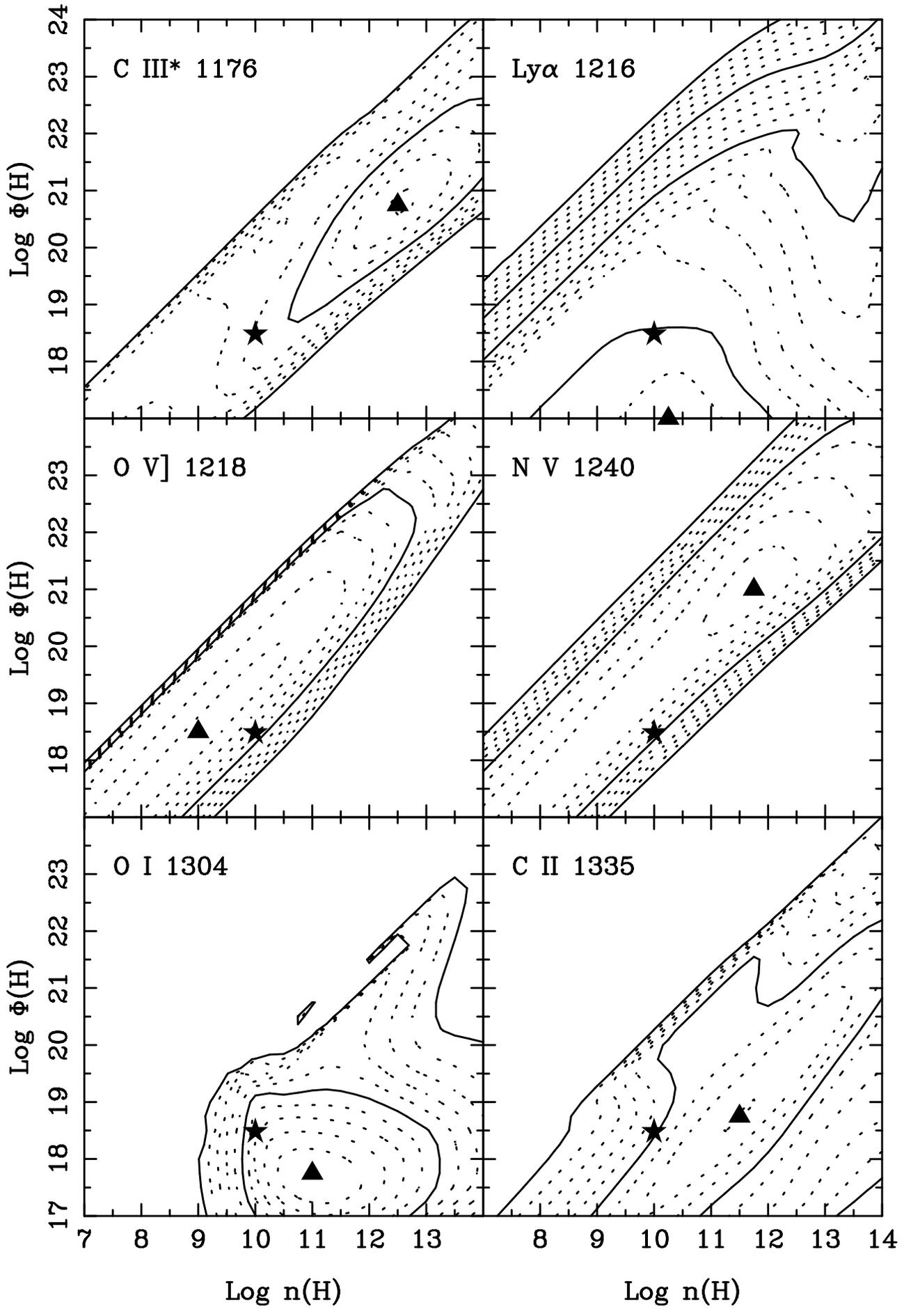
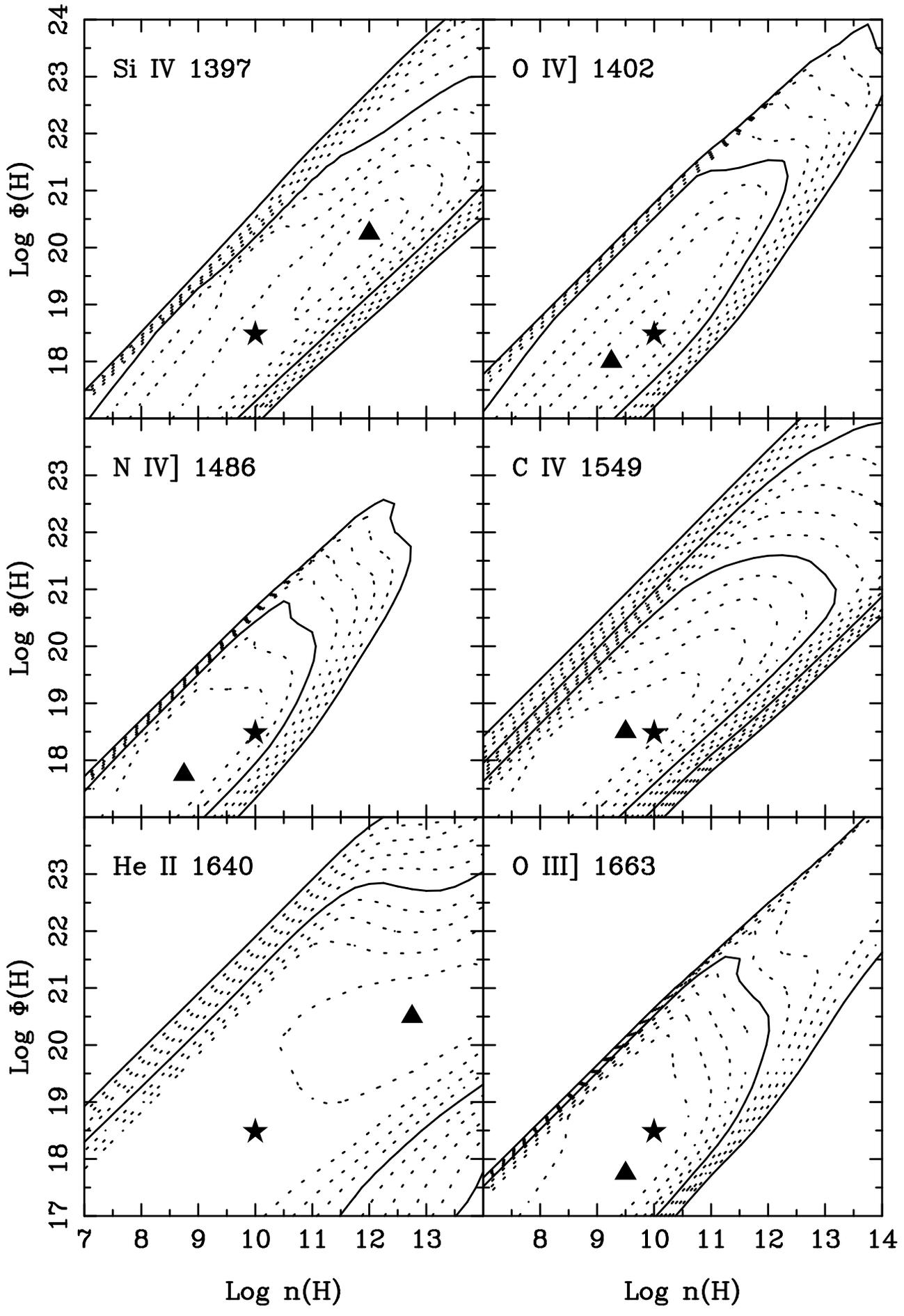
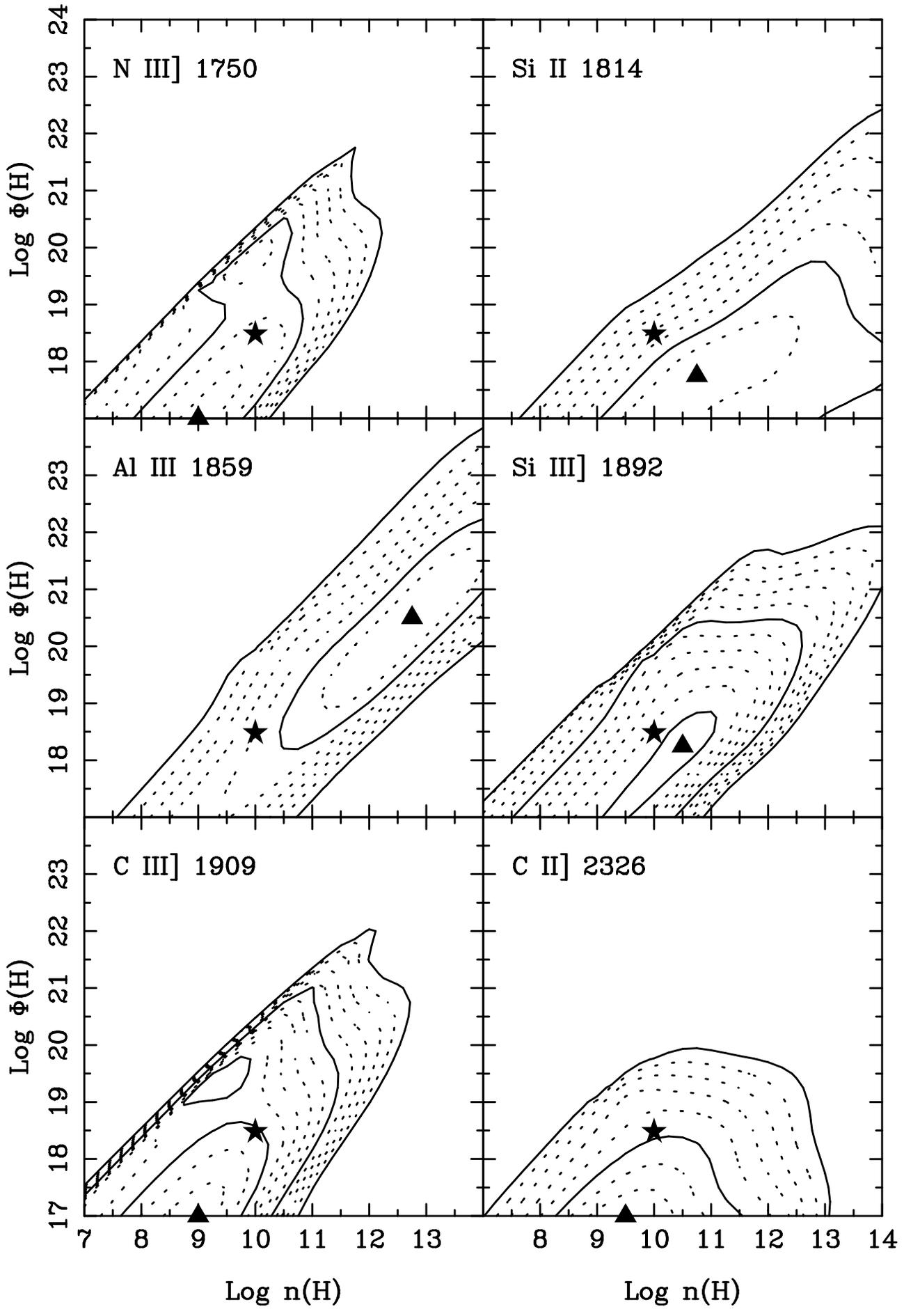
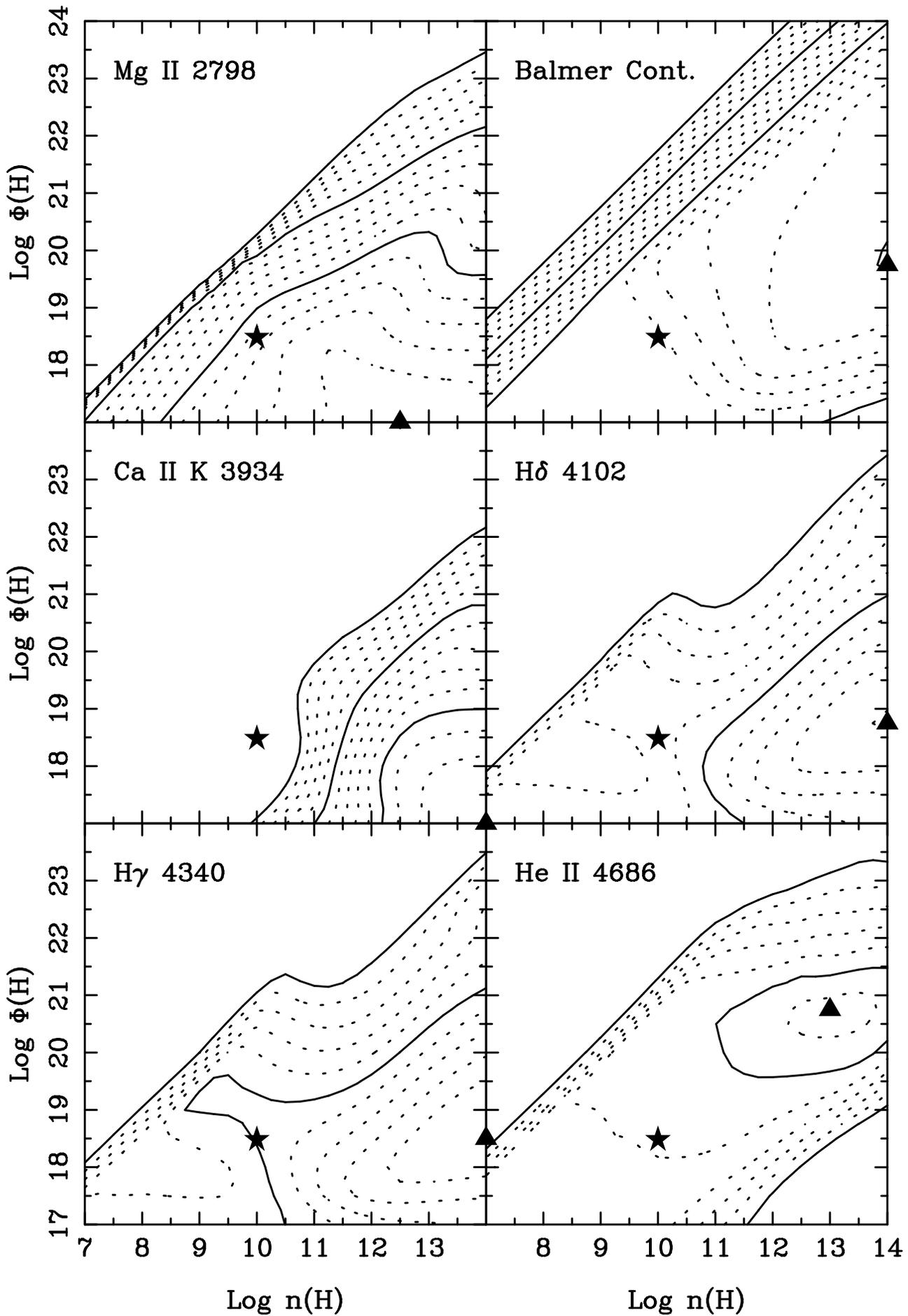
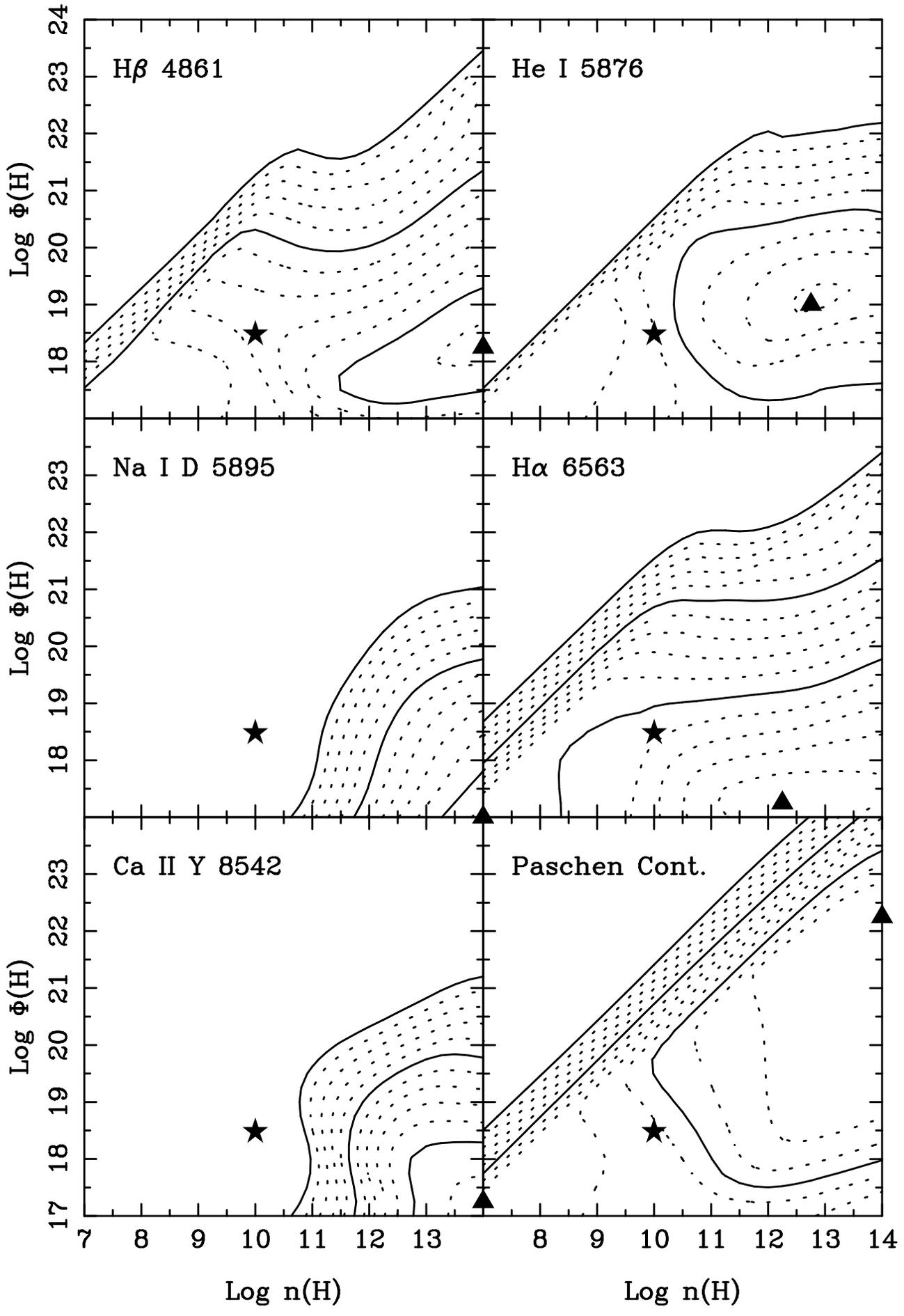
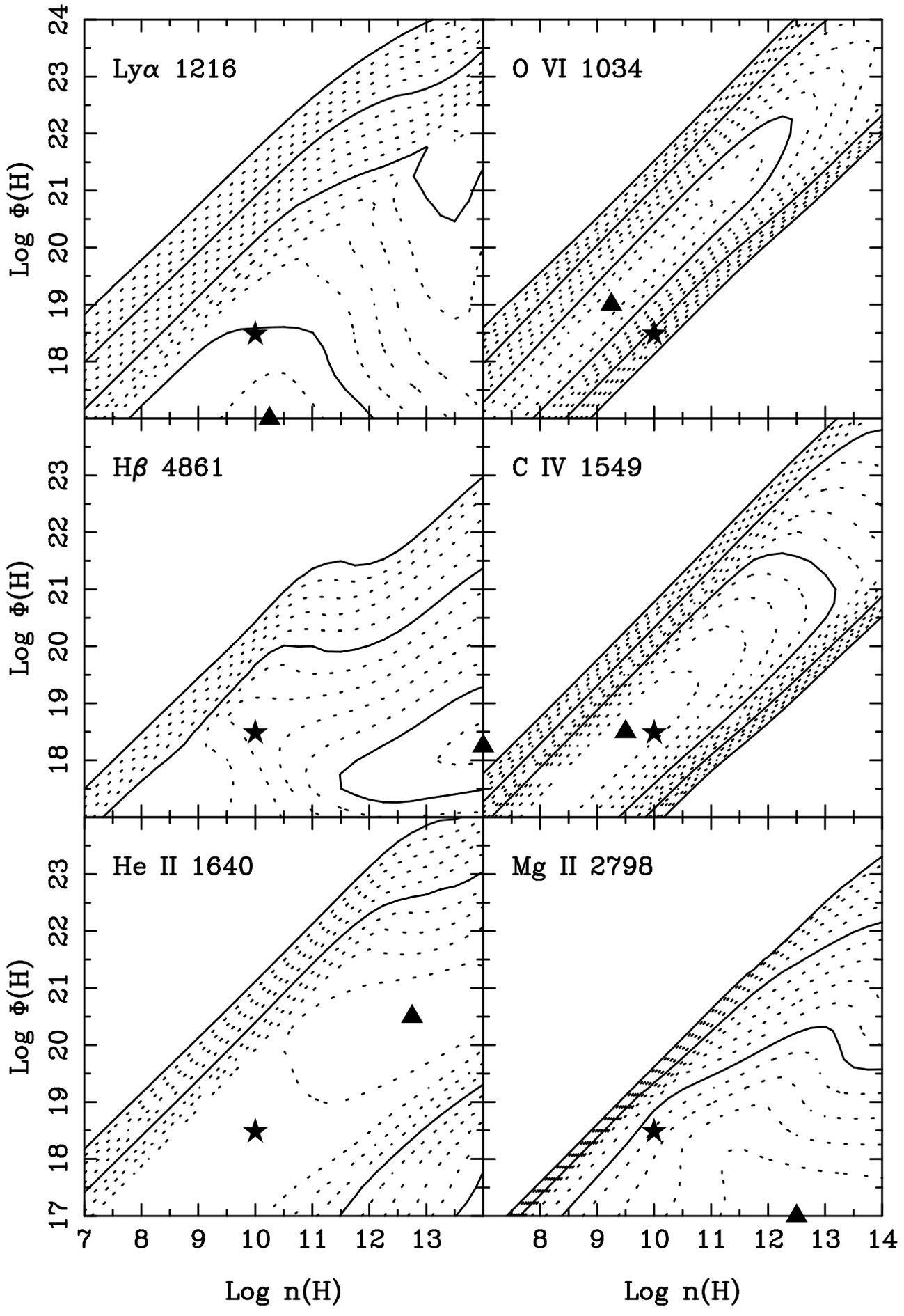
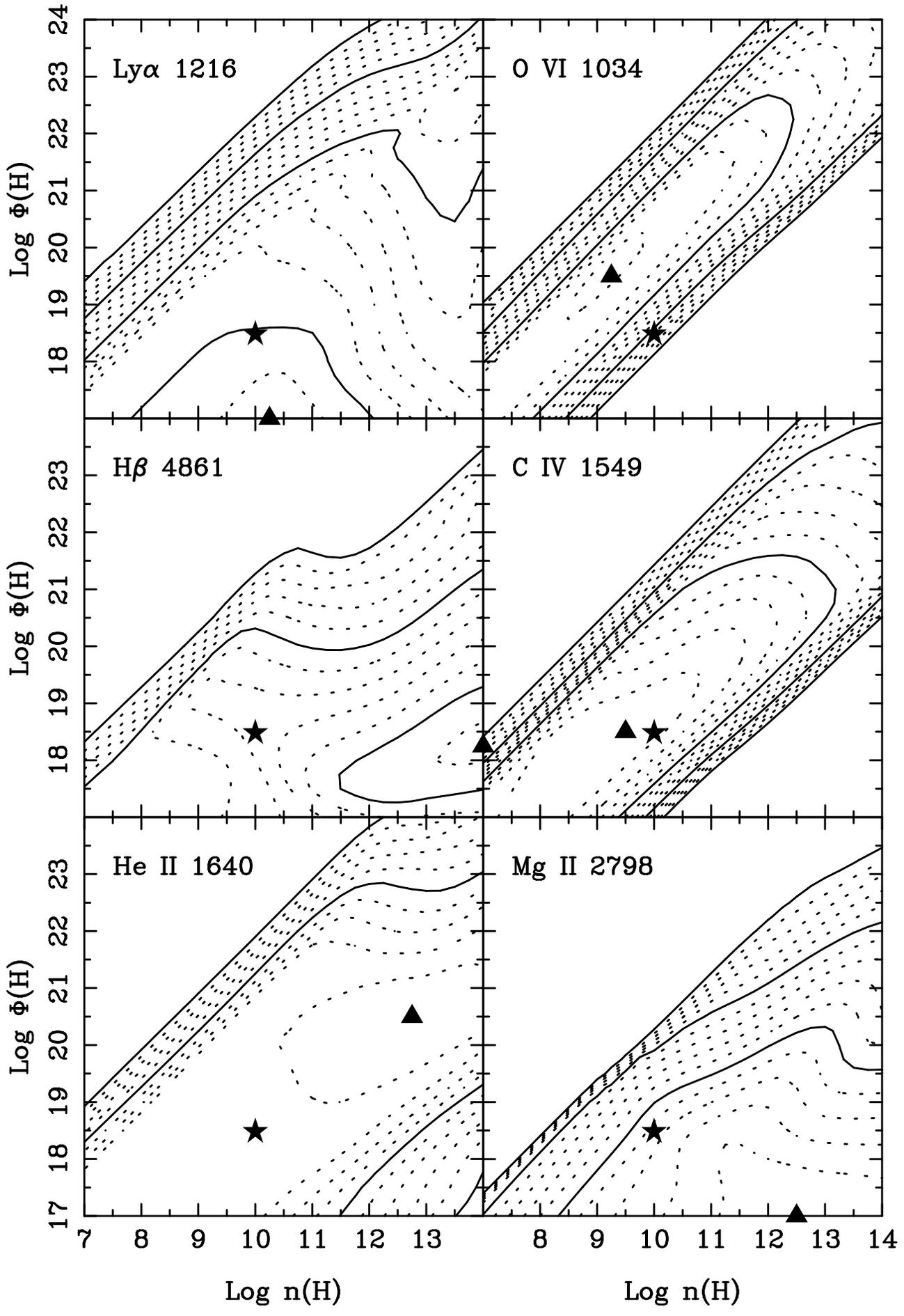
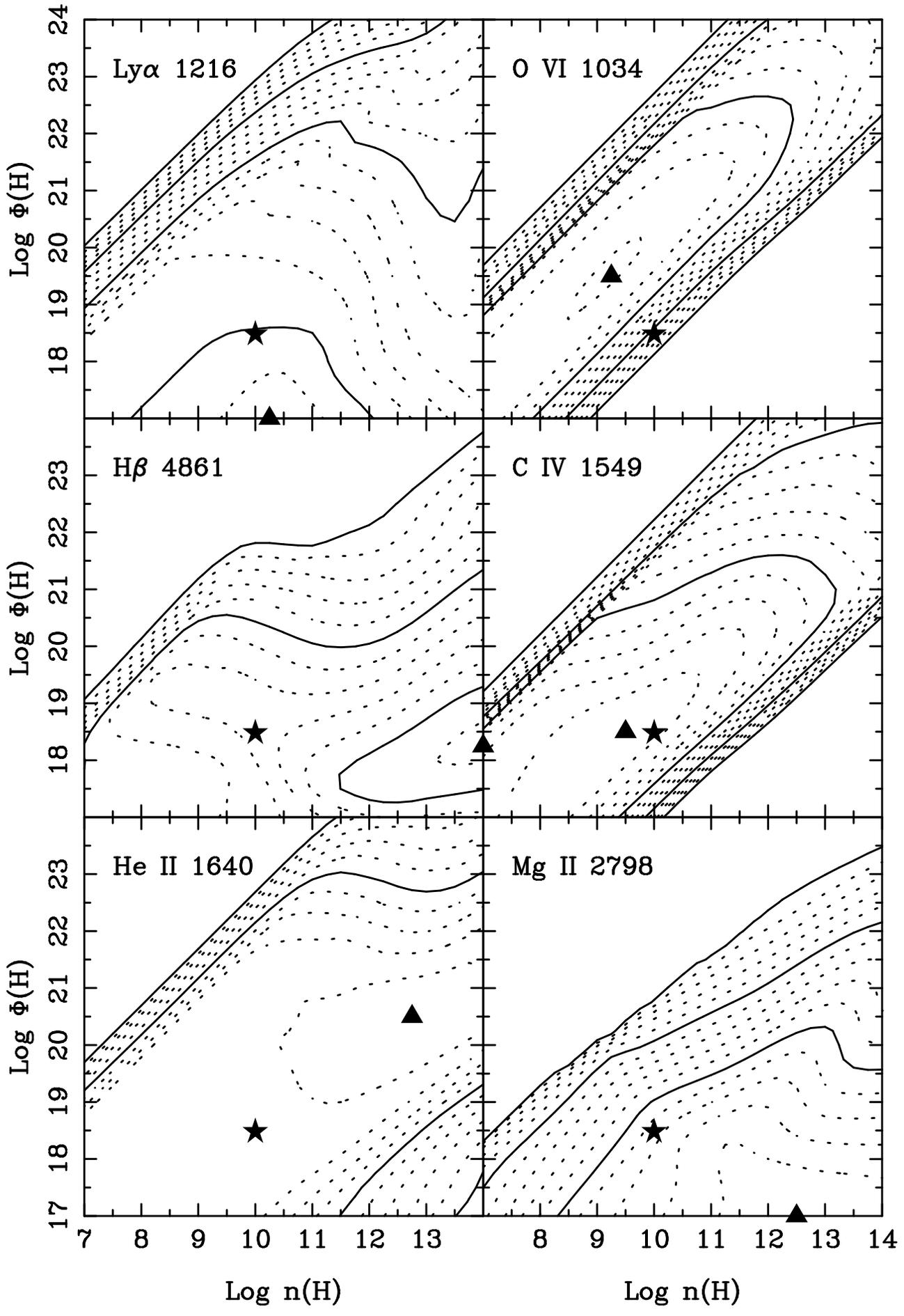
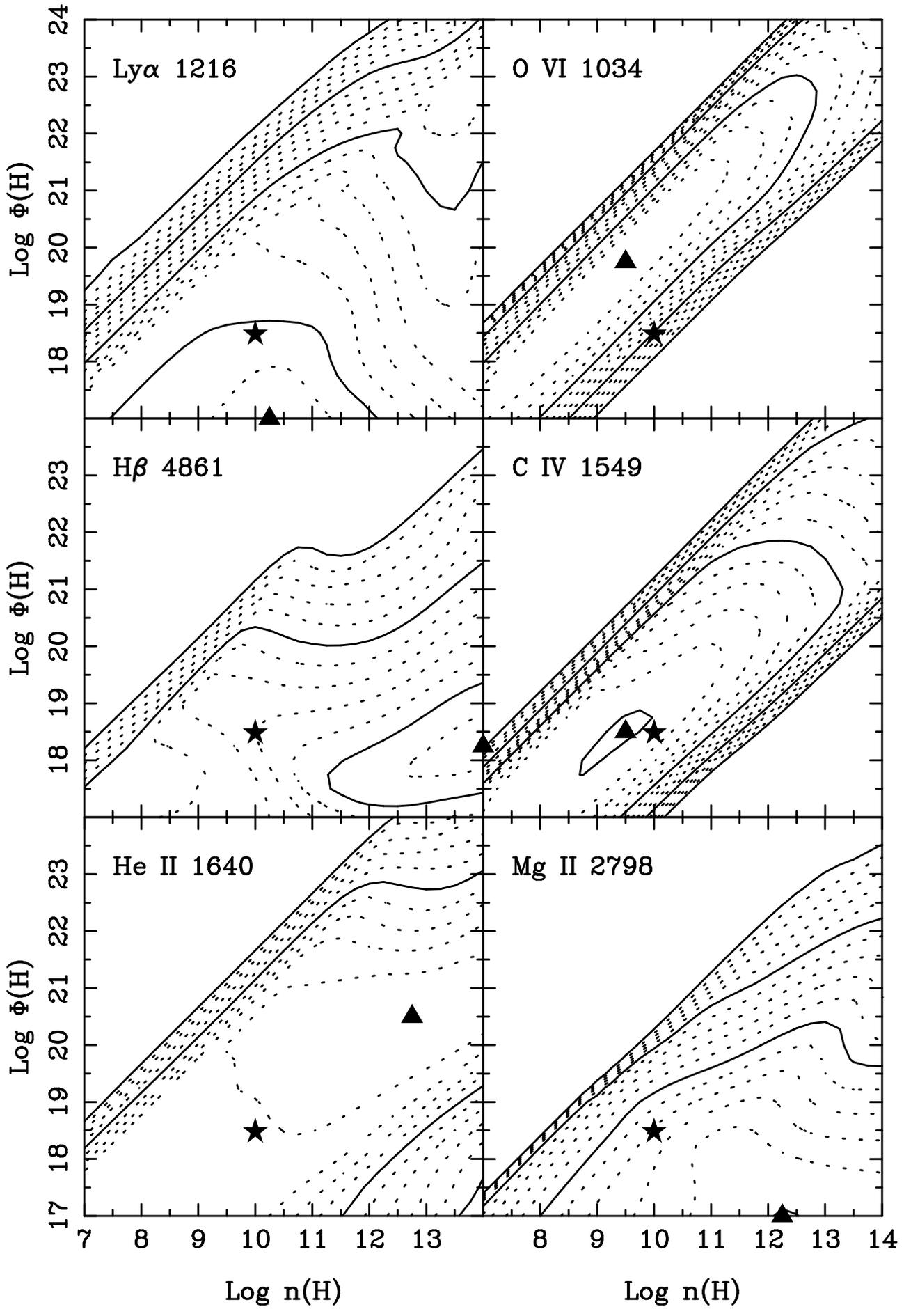
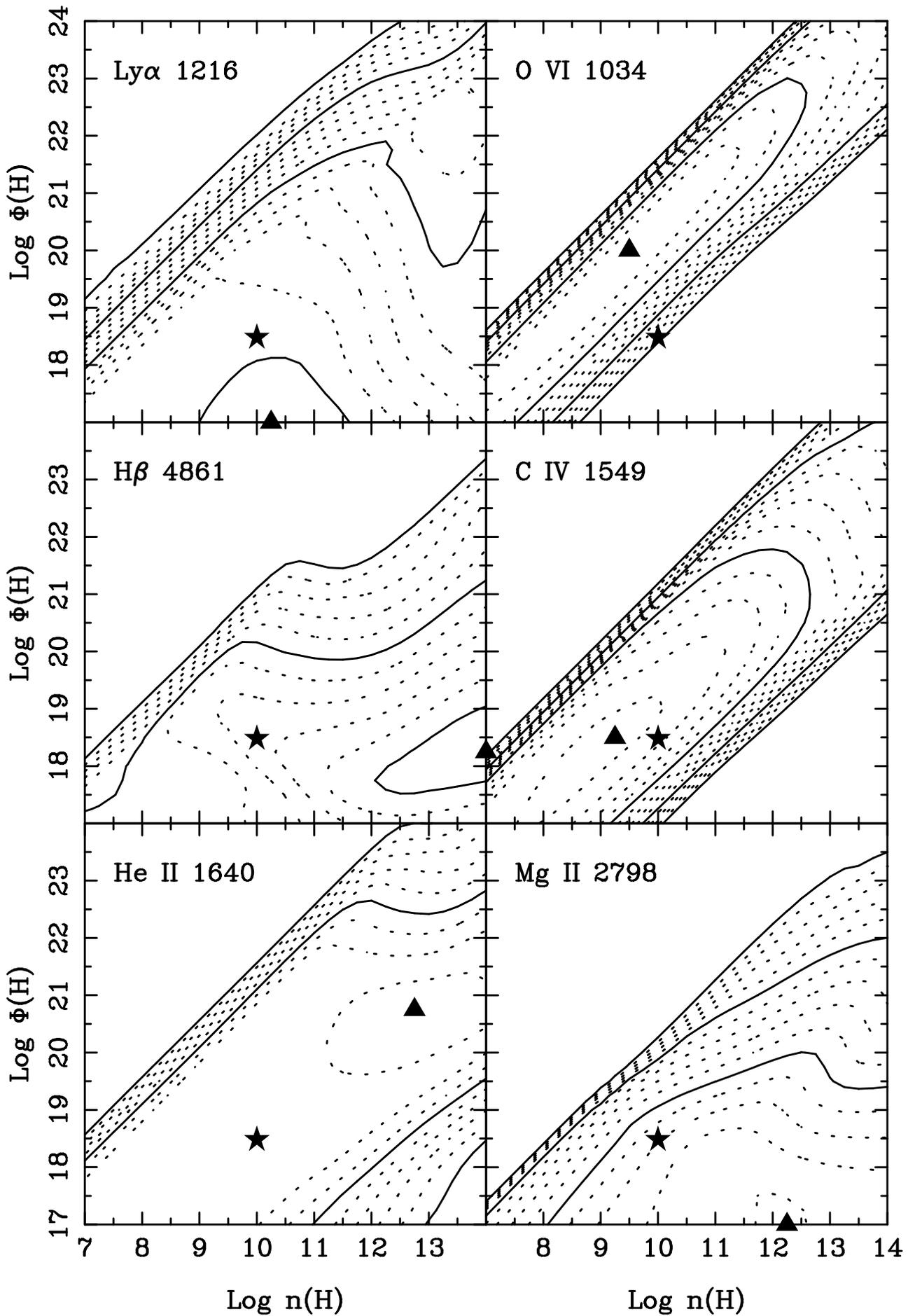
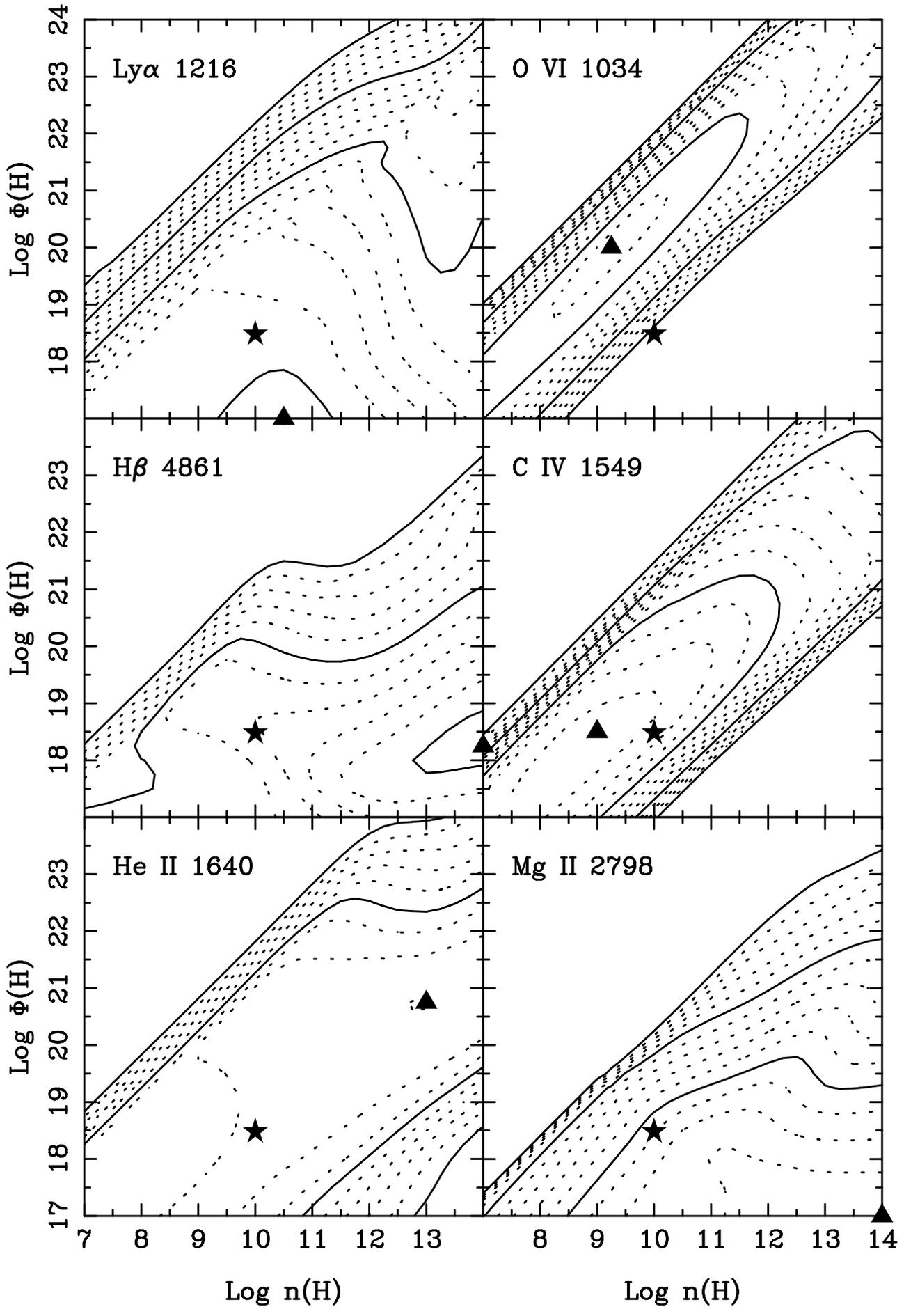
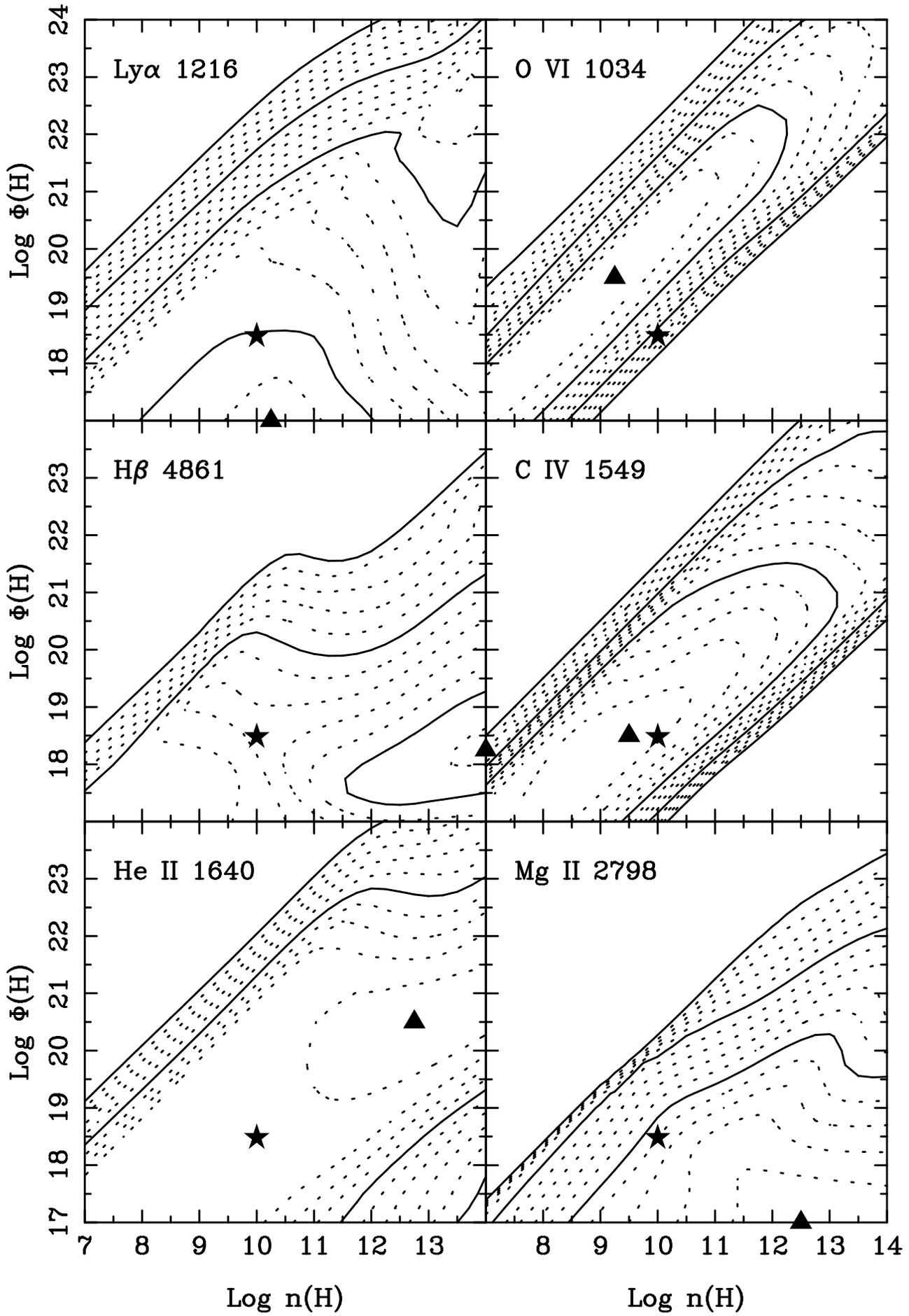
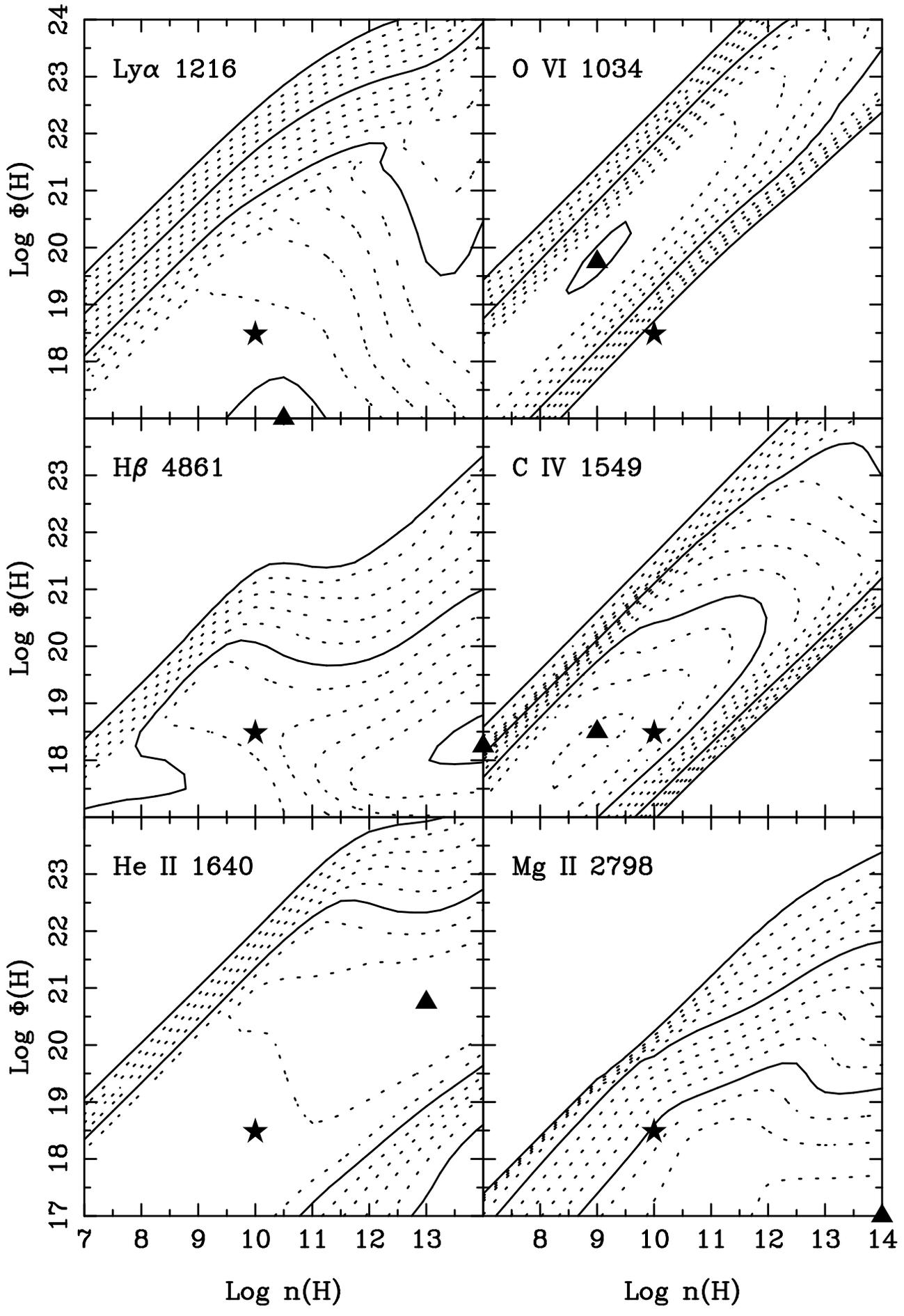
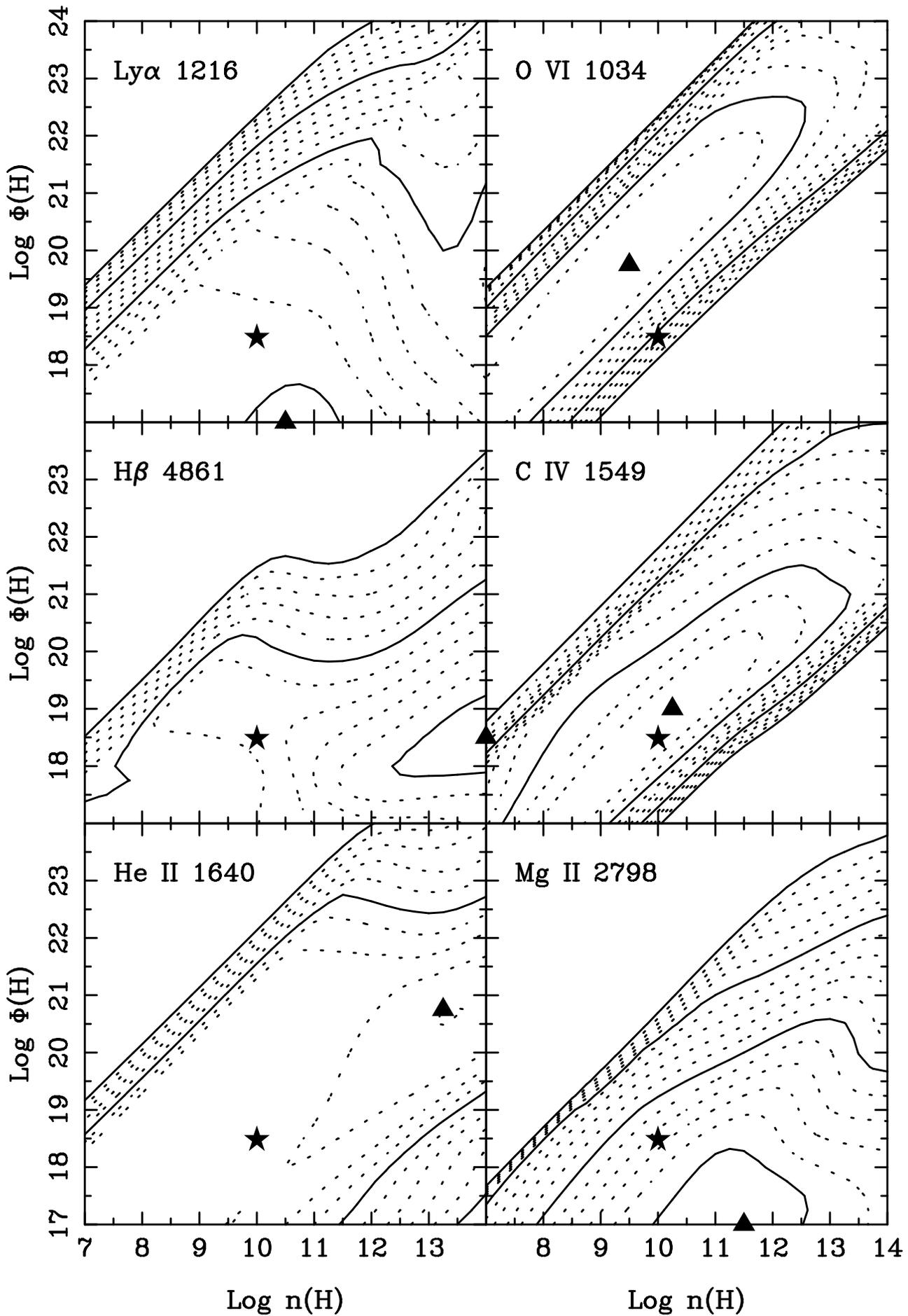
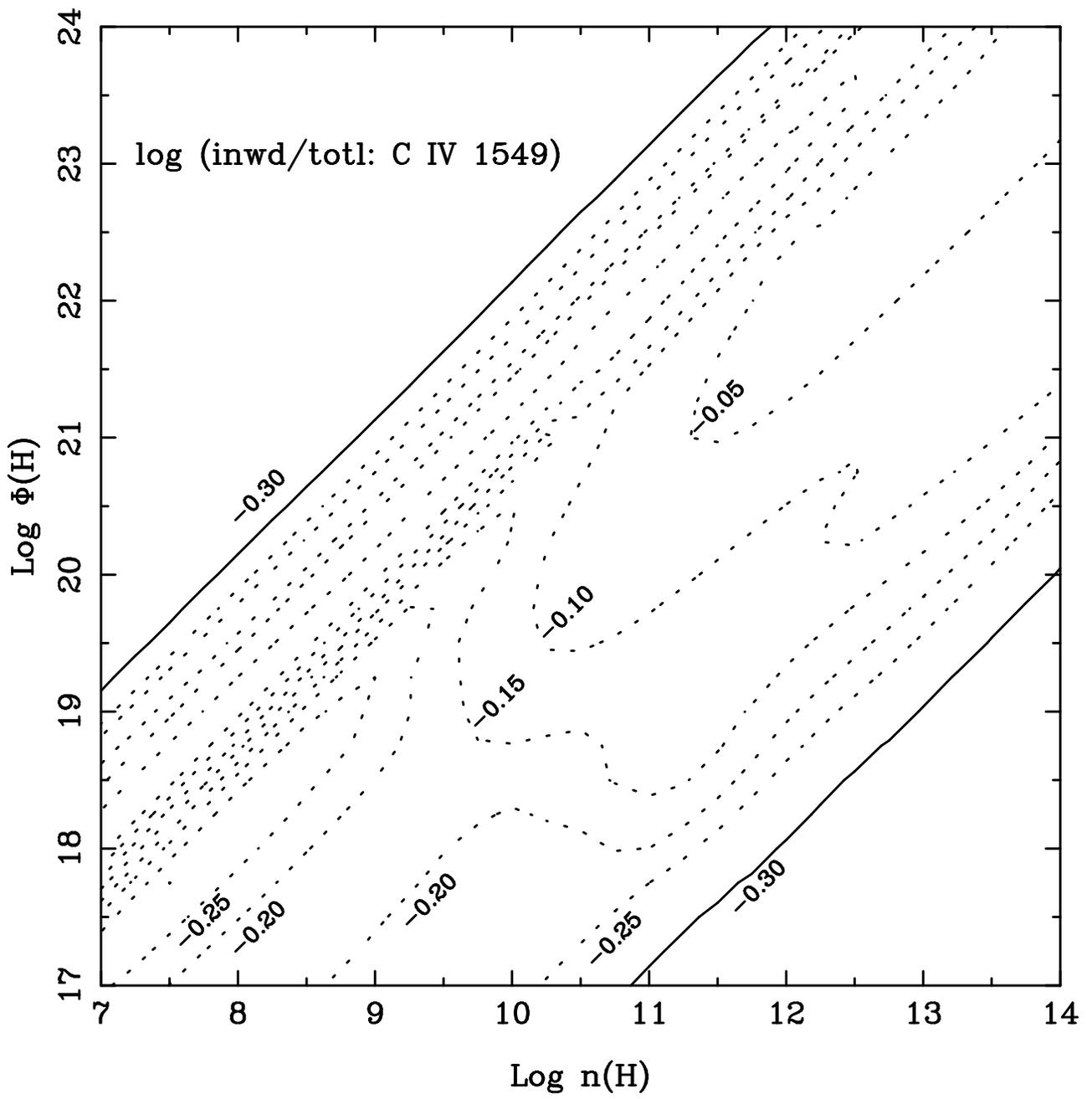
\end{document}